\documentclass[preprint2]{aastex6}
\usepackage{amsfonts,amsmath,amssymb,ulem,nicefrac,verbatim,color}
\usepackage{array,graphicx}
\usepackage{affils}

\def\simlt{\mathrel{\hbox{\rlap{\hbox{\lower4pt\hbox{$\sim$}}}\hbox{$<$}}}}
\def\simgt{\mathrel{\hbox{\rlap{\hbox{\lower4pt\hbox{$\sim$}}}\hbox{$>$}}}}

\def\ale{\mathrel{\hbox{\rlap{\hbox{\lower4pt\hbox{$\sim$}}}\hbox{$<$}}}}
\def\age{\mathrel{\hbox{\rlap{\hbox{\lower4pt\hbox{$\sim$}}}\hbox{$>$}}}}

\def\spose#1{\hbox to 0pt{#1\hss}}

\begin{document}

\title{A Reverse Shock and Unusual Radio Properties in GRB\,160625B}

\DeclareAffil{cfa}{Harvard-Smithsonian Center for Astrophysics, 60 Garden St., Cambridge, MA 02138, USA}
\DeclareAffil{nrao}{National Radio Astronomy Observatory, 520 Edgemont Road, Charlottesville, VA 22903, USA}
\DeclareAffil{UCB}{Department of Astronomy, University of California, 501 Campbell Hall, Berkeley, CA 94720-3411, USA}
\DeclareAffil{az}{Steward Observatory, University of Arizona, 933 N. Cherry Avenue, Tucson, AZ 85721, USA}
\DeclareAffil{ef}{Einstein Fellow}
\DeclareAffil{bath}{Department of Physics, University of Bath, Claverton Down, Bath, BA2 7AY, UK}
\DeclareAffil{slov}{Faculty  of  Mathematics  and  Physics,  University  of  Ljubljana, Jadranska 19, 1000 Ljubljana, Slovenia}
\DeclareAffil{liver}{Astrophysics Research Institute, Liverpool John Moores University, 
IC2, Liverpool Science Park, 146 Brownlow Hill, Liverpool L3 5RF, UK}
\DeclareAffil{slov2}{Centre for Astrophysics and Cosmology, University of Nova Gorica, Vipavska 11c, 5270 Ajdov\v s\v cina, Slovenia}
\DeclareAffil{mex}{Instituto de Astronom\'ia, Universidad Nacional Aut\'onoma de M\'exico, Ciudad
de M\'exico, M\'exico}
\DeclareAffil{it}{Dept. Physics and Earth Science, University of Ferrara, via Saragat 1, I-44122, Ferrara, Italy}
\DeclareAffil{leic}{University of Leicester, Department of Physics \& Astronomy and Leicester Institute of Space \& Earth Observation, University Road, Leicester, LE1 7RH, UK}

\affilauthorlist{K.~D.~Alexander\affils{cfa}, T.~Laskar\affils{nrao,UCB}, E.~Berger\affils{cfa}, C.~Guidorzi\affils{it}, S.~Dichiara\affils{mex}, W.~Fong\affils{az,ef}, A.~Gomboc\affils{slov2}, S.~Kobayashi\affils{liver}, D.~Kopac\affils{slov}, C.~G.~Mundell\affils{bath}, N.~R.~Tanvir\affils{leic}, P.~K.~G.~Williams\affils{cfa}}

\begin{abstract}
We present multi-wavelength observations and modeling of the exceptionally bright long $\gamma$-ray burst GRB 160625B. The optical and X-ray data are well-fit by synchrotron emission from a collimated blastwave with an opening angle of $\theta_j\approx 3.6^\circ$ and kinetic energy of $E_K\approx 2\times10^{51}$ erg, propagating into a low density ($n\approx 5\times10^{-5}$ cm$^{-3}$) medium with a uniform profile. The forward shock is sub-dominant in the radio band; instead, the radio emission is dominated by two additional components. The first component is consistent with emission from a reverse shock, indicating an initial Lorentz factor of $\Gamma_0\gtrsim 100$ and an ejecta magnetization of $R_B\approx 1-100$. The second component exhibits peculiar spectral and temporal evolution and is most likely the result of scattering of the radio emission by the turbulent Milky Way interstellar medium (ISM). Such scattering is expected in any sufficiently compact extragalactic source and has been seen in GRBs before, but the large amplitude and long duration of the variability seen here are qualitatively more similar to extreme scattering events previously observed in quasars, rather than normal interstellar scintillation effects. High-cadence, broadband radio observations of future GRBs are needed to fully characterize such effects, which can sensitively probe the properties of the ISM and must be taken into account before variability intrinsic to the GRB can be interpreted correctly. 
\smallskip
\end{abstract}


\section{Introduction}

Long duration $\gamma$-ray bursts (GRBs) have been conclusively linked to the collapse of massive stars \citep{wb06}, but many questions about their progenitors and the physics powering GRB jets remain. The jet's composition and initial Lorentz factor can be probed directly through observations of synchrotron emission from the reverse shock (RS), produced when the jet begins to interact with the circumburst medium \citep{mr93,sp99}. Strong RS signatures are predicted when the energy density of the jet is dominated by baryons, while a weaker or absent RS may indicate a jet dominated by Poynting flux \citep{sp99}. RS emission fades quickly and later emission is dominated by the forward shock (FS) between the ejecta and the surrounding material \citep{spn98,sp99}, making early observations  essential to constrain RS models.

The brightest RS signature is predicted in the optical band on $\lesssim$ hour timescales, but despite early optical observations enabled by robotic telescopes and rapid X-ray and ultraviolet (UV) localizations of GRBs by {\it Swift}, to date only a small fraction of GRBs exhibit unambiguous optical RS signatures (\citealt{jap14} and references therein.) Bright optical flashes are now ruled out by observations in many events, while other events show complicated optical light curves that, like the prompt $\gamma$-ray emission, may originate instead from internal shocks \citep{kop13,jap14}. Some authors have proposed that RS emission may be easier to observe at longer wavelengths, where the emission peaks on timescales of days \citep{mun07,mel10,kop15}. This approach was successfully adopted in trailblazing multi-frequency radio studies of GRB 130427A that characterized the RS emission at multiple epochs in detail \citep{lbz+13, per14}. In 2015, we began an intensive observing campaign at the Karl G. Jansky Very Large Array (VLA) to obtain additional early radio observations of long GRBs, resulting in a second multi-frequency detection of RS emission in GRB 160509A \citep{lab+16}. 

Here, we present new results from our VLA campaign for the {\it Fermi} GRB 160625B. We combine our detailed multi-frequency radio observations with optical and X-ray data, using a full MCMC statistical analysis to constrain the burst properties. The radio emission is dominated by a bright RS at early times and exhibits additional strong variability at late times, plausibly due to scattering by structures in the Galactic interstellar medium along the line of sight. All errorbars are $1\sigma$ confidence intervals unless otherwise stated and all magnitudes are in the AB system \citep{oke83}. We assume an event redshift of $z=1.406$ (determined from optical spectroscopy of the afterglow; \citealt{gcn600}) and standard $\Lambda$CDM cosmology with $\Omega_m=0.27$, $\Omega_{\Lambda}=0.73$, and $H_0=71$ km s$^{-1}$ Mpc$^{-1}$ throughout.

\section{GRB Properties and Observations}

\subsection{$\gamma$-rays}\label{sec:he}

GRB 160625B was discovered by the {\it Fermi} Gamma-ray Space Telescope on 2016 June 25 \citep{drm+16}. The burst triggered the Gamma-ray Burst Monitor (GBM; \citealt{mee09}) at 22:40:16.28 UTC and 22:51:16.03 UTC, and the Large Area Telescope (LAT; \citealt{at09}) at 22:43:24.82 UTC \citep{b16}. The burst was also detected by Konus-Wind, Integral, and CALET. The initial GBM trigger was a soft peak with a duration of $T_{90}=0.84$ s and a fluence of $(1.75\pm0.05)\times10^{-6}$ erg cm$^{-2}$ (8 keV $-$ 40 MeV). This precursor was followed by $\sim180$ s of quiescence and then by the main emission episode, which was extremely bright and had a duration of $T_{90}=35$ s and a fluence of $(6.01\pm0.02)\times10^{-4}$ erg cm$^{-2}$. A third period of weak emission with a duration of $T_{90}=212$ s and a fluence of $(5.65\pm0.02)\times10^{-5}$ erg cm$^{-2}$ followed after another $\sim339$ s gap \citep{zzc+16}. For our analysis, we take $t_0$ to be the time of the LAT trigger, which coincides with the onset of the main emission episode, and take $T_{90}=35$ s for the burst because this episode comprises $>90$\% of the high-energy emission. The total isotropic-equivalent energy of the prompt emission is $E_{\gamma,{\rm iso}}\approx3\times10^{54}$ erg \citep{zzc+16}. The prompt emission is discussed in detail in \cite{zzc+16}, \cite{wang17}, and \cite{llz+17}.

\subsection{X-ray: \textit{Swift}/XRT}
\begin{figure*} 
\centerline{\includegraphics[width=5.2in]{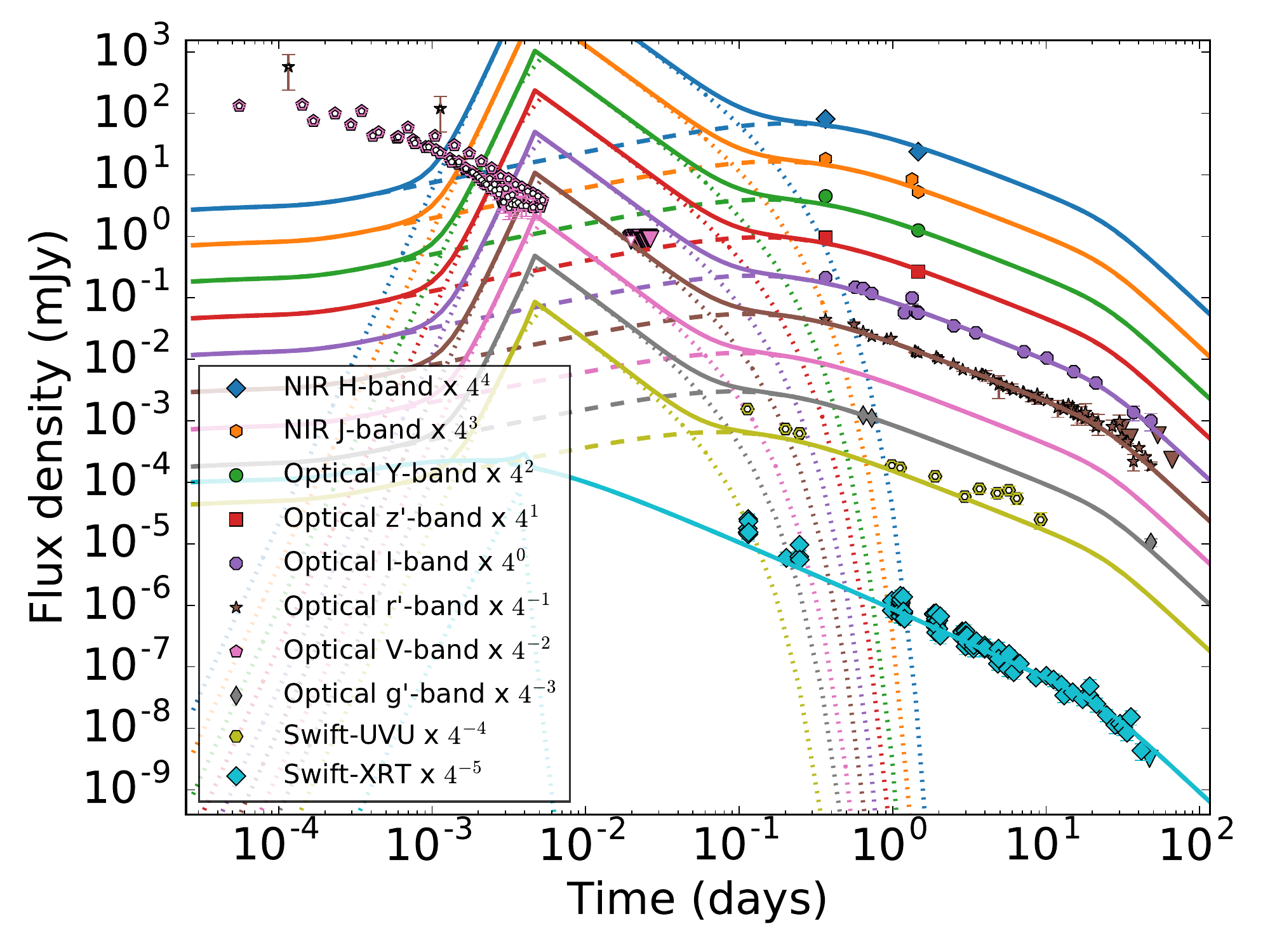}}
\centerline{\includegraphics[width=5.2in]{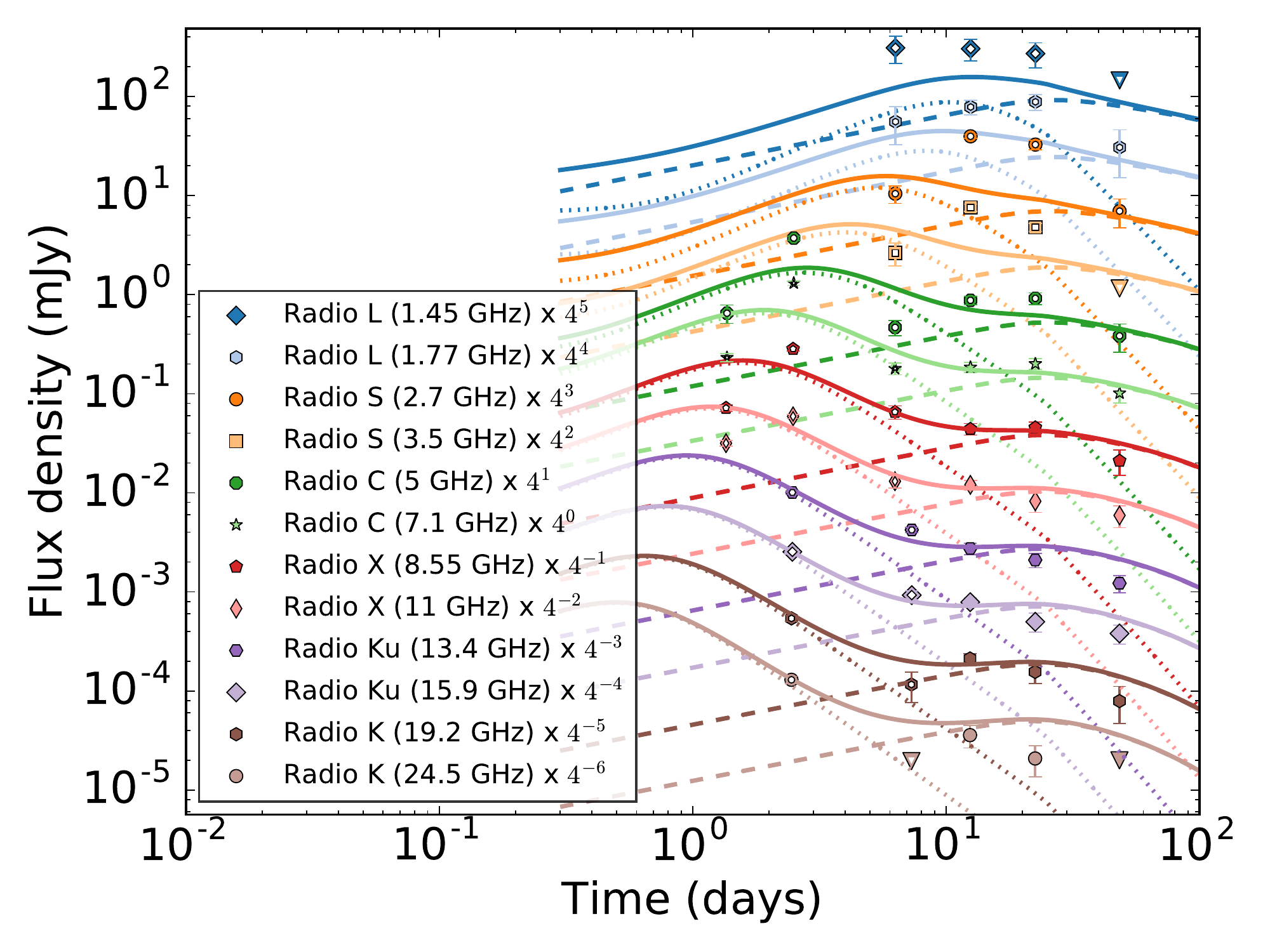}} 
\caption{Light curves of GRB 160625B, vertically shifted for clarity. We take $t=0$ to be the LAT trigger time. The best-fit model (solid lines; Table \ref{tab:mod}) consists of a forward shock (dashed component) and a Newtonian reverse shock (dotted component; Model 1). The optical and X-ray data drive the properties of the forward shock (top), while the reverse shock dominates the radio emission at early times (bottom). The optical detections before 0.01 d are likely related to the prompt emission, consistent with the sub-dominant extrapolated flux of the reverse shock at early times. These early data are excluded from our model fitting, as is the portion of the radio emission showing evidence of multiple components. The excluded points are indicated with open symbols.}
\label{fig:lc}
\end{figure*}

The {\it Swift} X-Ray Telescope (XRT; \citealt{gcg+04}) began tiled observations of the {\it Fermi} error circle 2.5 h after the trigger and at 2.7 h detected a bright, uncatalogued X-ray source determined to be the afterglow \citep{mel16}. XRT continued to observe the afterglow for 47 days, with the last detection at 41.7 days\footnote{\url{http://www.swift.ac.uk/xrt_live_cat/00020667/}}. There are two breaks in the count-rate light curve, at $t_1\approx1.23\times10^4$ s and $t_2\approx1.8\times10^6$ s. The intervals $t<t_1$ and $t>t_2$ do not contain sufficient data to construct spectra with high enough signal-to-noise to rule out spectral evolution across the breaks, so we exclude these time ranges from our spectral analysis. We use the online tool from the {\it Swift} website \citep{ev07,ev09} to extract a PC-mode spectrum from the time interval $t_1 < t < t_2$ and fit the spectrum with a photoelectrically absorbed power-law model with the Galactic neutral hydrogen column fixed to $N_{\rm H,MW} = 9.76\times10^{20}$ cm$^{-2}$ \citep{wil13}. We determine the photon index to be $\Gamma_{\rm X} = 1.86^{+0.10}_{-0.09}$ and the intrinsic absorption in the host galaxy to be $N_{\rm H,int} = 2.1^{+1.9}_{-1.8} \times 10^{21}$ cm$^{-2}$, with 90\% confidence. $N_{\rm H,int}$ is consistent with zero at the $\sim2\sigma$ level, but we keep $N_{\rm H,int} = 2.1 \times 10^{21}$ cm$^{-2}$ when computing the counts-to-flux ratio. We use the corresponding spectral index $\beta_{\rm X}=1-\Gamma_{\rm X}=-0.86_{-0.10}^{+0.09}$ and the associated counts-to-absorbed flux ratio of $3.6 \times 10^{-11}$ erg cm$^{-2}$ ct$^{-1}$ to convert the count rate to the observed flux density at 1 keV. The X-ray light curve is shown in Figure \ref{fig:lc}. 

\subsection{UV/Optical: \textit{Swift}/UVOT}\label{sec:uvot}
The {\it Swift} UV/Optical Telescope (UVOT; \citealt{rom05}) began observing the burst 2.7 h after the {\it Fermi} trigger, detecting a bright source in $U$ band \citep{oates16}. Additional observations were conducted in the $U$, $W1$, $M2$, and $V$ filters. The photometry was complicated by the presence of a nearby bright star, which created reflections that dominated the counts at the source position in many images and rendered the bluer bands entirely unusable. We restrict our analysis to the $U$ band images, where the source is clearly detected and the background is more uniform. 

We analyze the $U$ band data using {\tt HEASoft} (v. 6.16). We perform photometry with a $5\arcsec$ aperture and a 15$\arcsec$ background region. We vary the position of the background region from image to image to avoid reflection artifacts from the nearby bright star and most closely match the background near the GRB, but caution that the flux errors thus obtained may be underestimated. Given the large systematic uncertainties, we do not include these data in our model fitting but they are shown for completeness in Figure \ref{fig:lc}. 

\medskip
\subsection{Optical/NIR: LCOGT, ORM, Magellan, GCN Circulars}\label{sec:opt}

We began observing GRB\,160625B with the 2-m Faulkes Telescope North (FTN), which is operated by Las Cumbres Observatory Global Network (LCOGT; \citealt{Brown13LCOGT}) on June 26.01 UT ($0.56$~days after the GRB) in the SDSS $r^{\prime}$ and $i^{\prime}$ filters. Observations with the FTN went on on a daily basis for almost a week, then the 2-m Liverpool Telescope (LT; \citealt{Steele04LT}) at the Observatorio del Roque de Los Muchachos (ORM) took over in the same filters with a cadence of a few days until 37~days post GRB. Bias and flat-field corrections were applied using the specific pipelines of the LCOGT and of the LT. The optical afterglow magnitudes were

\begin{center}
\setlength\LTcapwidth{4.5in}
\begin{longtable*}{ccccccc}
\caption{Optical Observations}
\label{tab:opt} \\
\hline
\hline\noalign{\smallskip}
$t$  & Obser- & Instru- & Filter & Magnitude & Frequency & Flux Density \\
(d)     & vatory  &  ment &  &  (AB) & ($10^{14}$ Hz)  & ($\mu$Jy)  {\smallskip} \\
\hline\noalign{}
0.56 & LCOGT & FTN & $r^{\prime}$ & $18.49 \pm 0.12$ & 4.56  & $146 \pm 17$ \\
0.57 & LCOGT & FTN & $i^{\prime}$ & $18.47 \pm 0.14$ & 3.93 & $150 \pm 20$ \\
1.19 & LCOGT & FTN & $i^{\prime}$ & $19.56 \pm 0.04$ & 3.93 & $57 \pm 12$ \\
1.40 & LCOGT & FTN & $r^{\prime}$ & $19.51 \pm 0.20$ & 4.56 & $60 \pm 3$ \\
1.41 & LCOGT & FTN & $i^{\prime}$ & $19.46 \pm 0.05$ & 3.93 & $60 \pm 3$ \\
1.42 & LCOGT & FTN & $r^{\prime}$ & $19.60 \pm 0.04$ & 4.56 & $53 \pm 3$ \\
1.46 & LCOGT & FTN & $i^{\prime}$ & $19.48 \pm 0.03$ & 3.93 & $59 \pm 3$ \\
2.49 & LCOGT & FTN & $r^{\prime}$ & $20.09 \pm 0.06$ & 4.56 & $33.4 \pm 1.9$ \\
2.50 & LCOGT & FTN & $i^{\prime}$ & $20.04 \pm 0.10$ & 3.93 & $35 \pm 3$ \\
3.47 & LCOGT & FTN & $r^{\prime}$ & $20.48 \pm 0.04$ & 4.56 & $23.3 \pm 1.2$ \\
3.49 & LCOGT & FTN & $i^{\prime}$ & $20.32 \pm 0.09$ & 3.93 & $27 \pm 2$ \\
4.54 & LCOGT & FTN & $r^{\prime}$ & $20.75 \pm 0.11$ & 4.56 & $18.2 \pm 1.9$ \\
5.52 & LCOGT & FTN & $r^{\prime}$ & $21.00 \pm 0.13$ & 4.56 & $14.5 \pm 1.8$  \\
7.17 & ORM & LT & $i^{\prime}$ & $21.09 \pm 0.03$ & 3.93 & $13.3 \pm 0.7$ \\
7.18 & ORM & LT & $r^{\prime}$ & $21.26 \pm 0.03$ & 4.56 & $11.4 \pm 0.6$ \\
10.12 & ORM & LT & $i^{\prime}$ & $21.35 \pm 0.03$ & 3.93 & $10.5 \pm 0.5$ \\
10.13 & ORM & LT & $r^{\prime}$ & $21.57 \pm 0.03$ & 4.56 & $8.6 \pm 0.4$ \\
15.13 & ORM & LT & $i^{\prime}$ & $21.9 \pm 0.08$ & 3.93 & $6.3 \pm 0.5$ \\
15.14 & ORM & LT & $r^{\prime}$ & $22.06 \pm 0.05$ & 4.56 & $5.5 \pm 0.3$ \\
21.09 & ORM & LT & $i^{\prime}$ & $22.36 \pm 0.10$ & 3.93 & $4.1 \pm 0.4$ \\
21.10 & ORM & LT & $r^{\prime}$ & $22.64 \pm 0.12$ & 4.56 & $3.2 \pm 0.4$ \\
37.10 & ORM & LT & $i^{\prime}$ & $23.56 \pm 0.26$ & 3.93 & $1.4 \pm 0.4$ \\
37.12 & ORM & LT & $r^{\prime}$ & $24.05 \pm 0.28$ & 4.56 & $0.9 \pm 0.3$ \\
48.13 & Magellan & LDSS3 & $i^{\prime}$ & $23.9 \pm 0.3$ & 3.93 & $1.0 \pm 0.3$ \\
48.15 & Magellan & LDSS3 & $r^{\prime}$ & $24.23 \pm 0.15$ & 4.56 & $0.74 \pm 0.11$ \\
48.18 & Magellan & LDSS3 & $g^{\prime}$ & $24.33 \pm 0.15$ & 6.29 & $0.67 \pm 0.10$ \\
\hline\noalign{\smallskip}
\caption[]{Optical observations of GRB 160625B from Las Cumbres Observatory (LCOGT), the Observatorio del Roque de Los Muchachos (ORM), and Magellan. All values of $t$ are relative to 2016 June 25 22:43:24.82 UT, the LAT trigger time. The data have not been corrected for extinction.}
\end{longtable*}
\end{center}

\vspace{-0.22in}
\noindent obtained by PSF-fitting photometry, after calibrating the zero-points with nine nearby stars with SDSS $r^{\prime}$ and $i^{\prime}$ magnitudes from the URAT1 catalog \citep{Zacharias15URAT1}. A systematic error of $0.02$~mag, due to the zero-point scatter of the calibrating stars, was added to the statistical uncertainties of magnitudes.

We subsequently observed GRB 160625B on 2016 August 12.12 UT (48.1 d after the burst) with LDSS-3 on the 6.5 m Magellan/Clay Telescope at Las Campanas Observatory. We obtained eight 180 s exposures in $i^{\prime}$ band, six 240 s exposures in $r^{\prime}$ band, and four 420 s exposures in $g^{\prime}$ band. The data were reduced using a custom IDL script and standard \texttt{IRAF} routines. The afterglow is detected in a stacked image in each filter. Aperture photometry was performed using nearby stars from the Pan-STARRS $3\pi$ survey \citep{pan16}.

Finally, we collected other optical and near-infrared (NIR) observations of GRB 160625B reported through the Gamma-ray Burst Coordinates Network (GCN) Circulars and by \cite{zzc+16} and converted all photometry to flux densities. These observations include early optical data from the Pi of the Sky North observatory \citep{bat16} and the Mini-MegaTORTORA telescope \citep{kar16,zzc+16}, which detected a bright optical flash coincident with the main peak of $\gamma$-ray emission. The flux densities derived from the Mini-MegaTORTORA photometry are systematically $\sim1.5$ times larger than flux densities from the simultaneous Pi of the Sky observations; this offset is due to either a calibration difference or the different filter bandpasses used by each instrument. Both groups used reference stars to perform a color correction and obtain approximate $V$ band magnitudes, but without a simultaneous spectrum an absolute photometric calibration is not possible. A precise calibration is not necessary for our results, as we only include these data in our modeling as an approximate upper limit on RS emission (Section \ref{sec:rs}). We list our Las Cumbres, ORM, and Magellan observations in Table \ref{tab:opt}. The fluxes reported in Table 1 have not been corrected for extinction, as this correction is included directly in our modeling framework (Section \ref{sec:mod}). We expect moderate Galactic extinction along the line of sight to the GRB: $A_g=0.42$, $A_r=0.29$, $A_i=0.22$, and $A_z=0.16$ \citep{schlaf11}. The optical light curves including all of the data used in our modeling are shown in Figure \ref{fig:lc}.

\subsection{Radio: VLA}
\begin{figure*} 
\centerline{\includegraphics[width=2.5in]{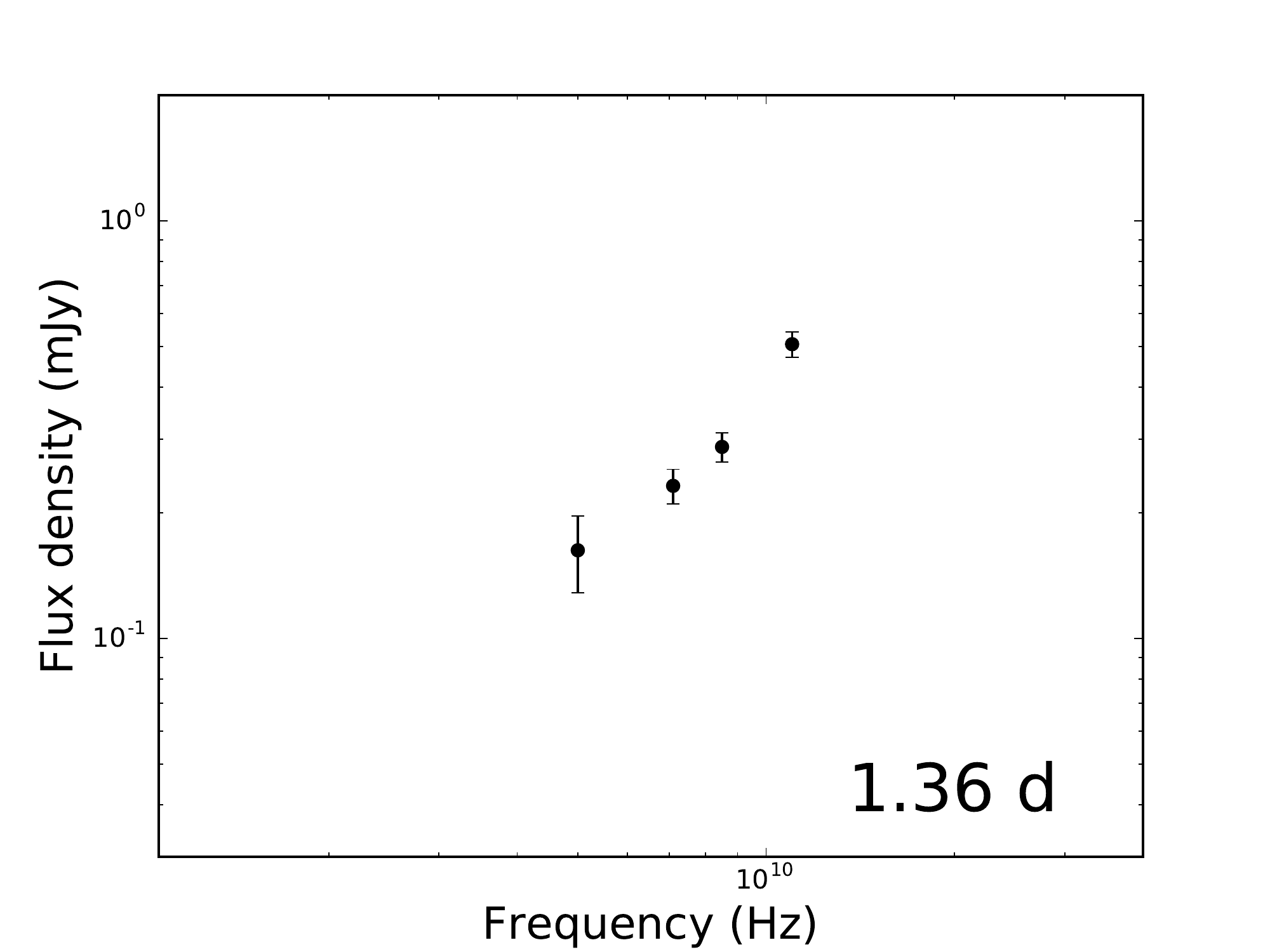}\includegraphics[width=2.5in]{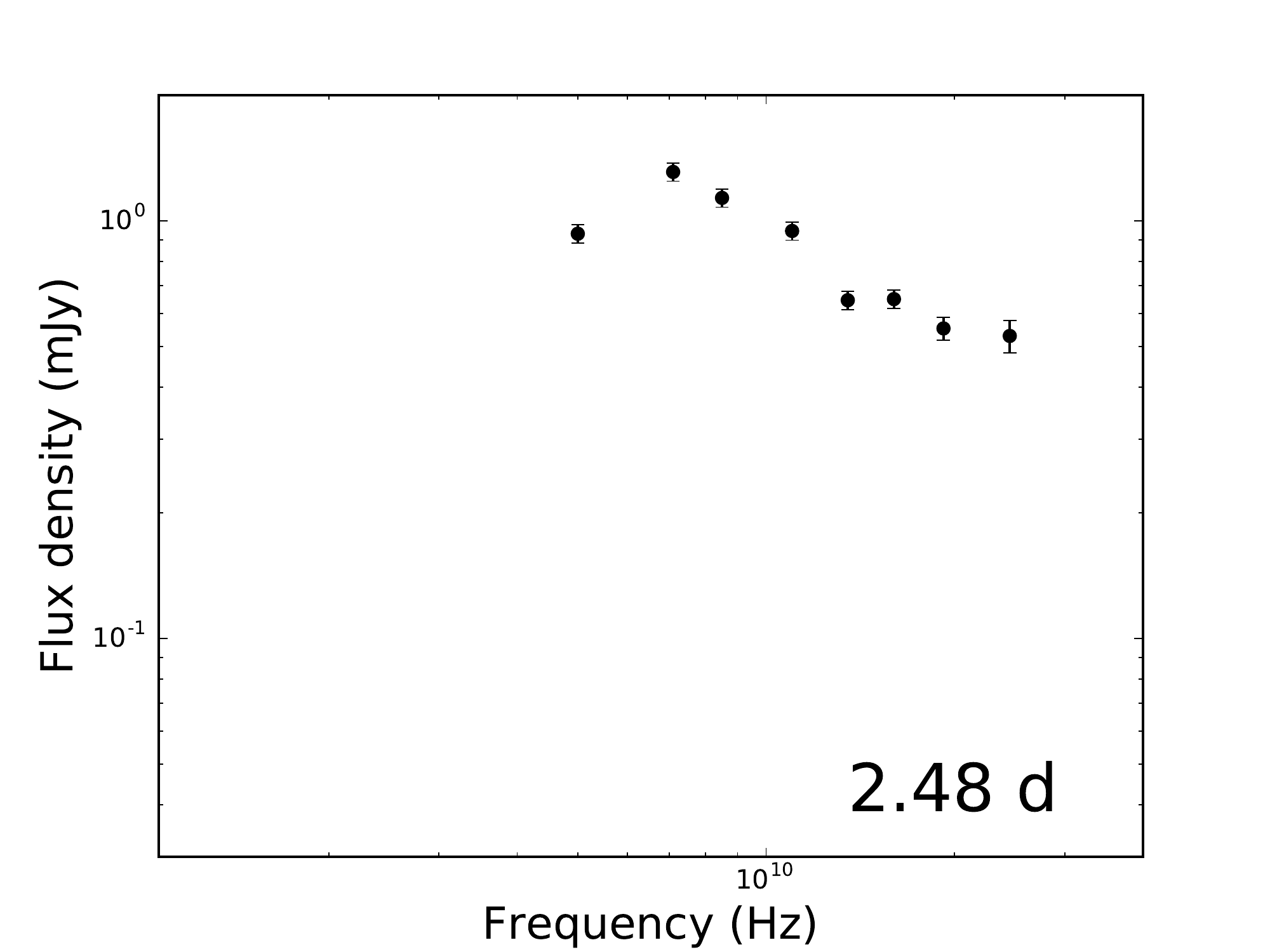}\includegraphics[width=2.5in]{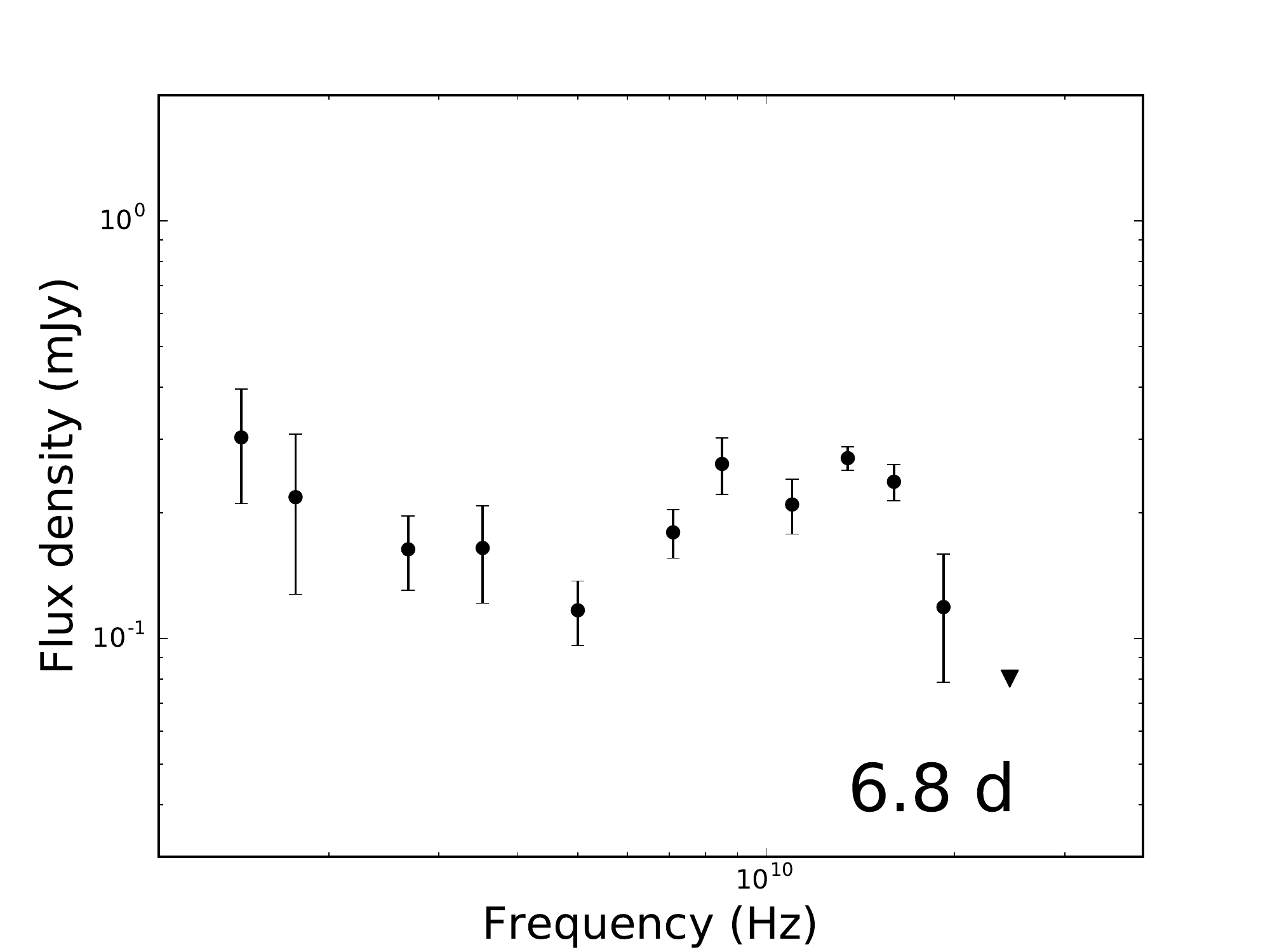}}
\centerline{\includegraphics[width=2.5in]{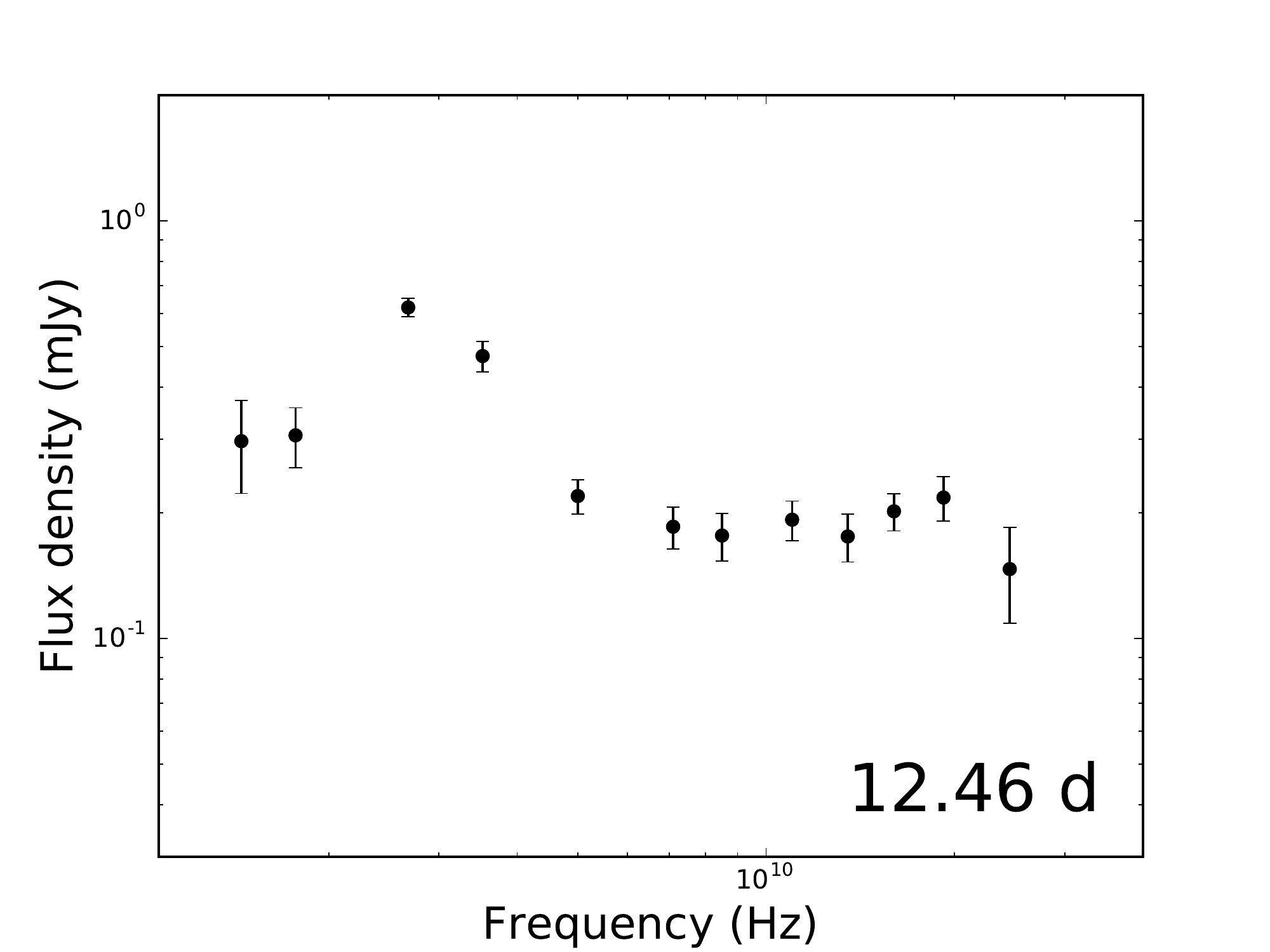}\includegraphics[width=2.5in]{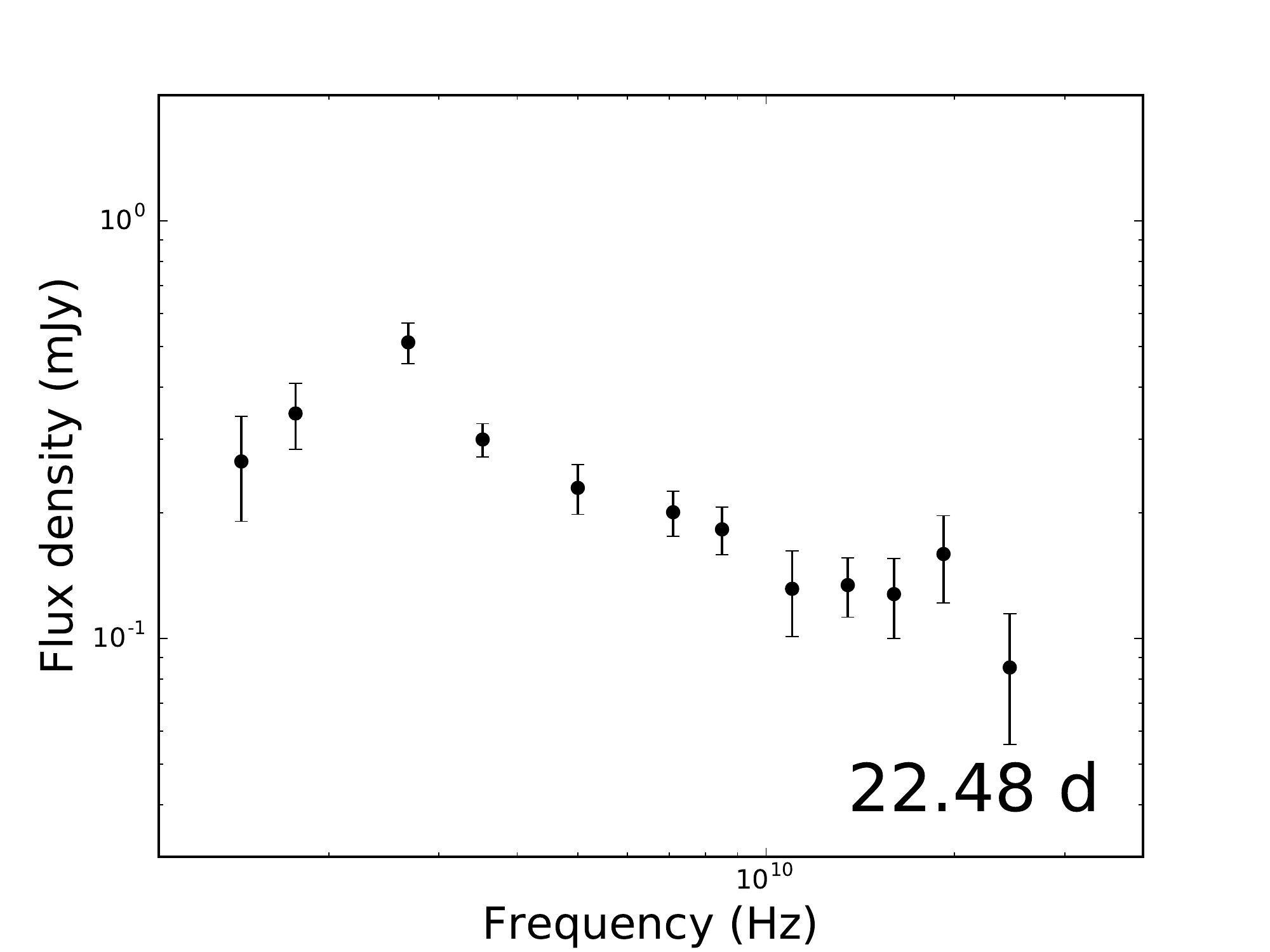}\includegraphics[width=2.5in]{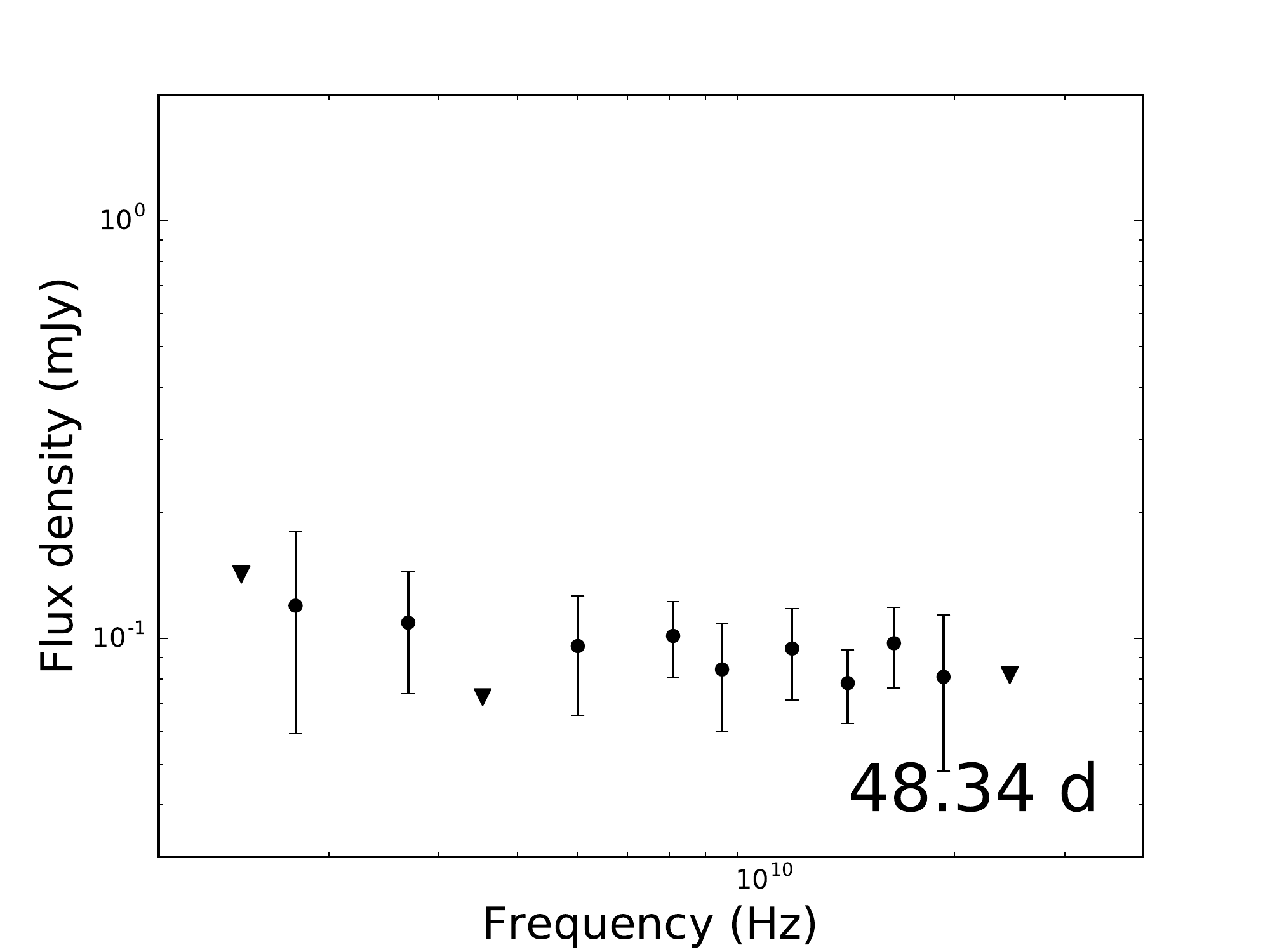}} 
\caption{Observed radio spectral energy distributions of GRB 160625B. The data show evidence of multiple components and the effects of interstellar scintillation.}
\label{fig:sed1}
\end{figure*}

We observed the afterglow using the Karl G. Jansky Very Large Array (VLA) starting 1.35 d after the burst. Our observations span frequencies between 1.45 GHz and 24.5 GHz and extend to 48.38 d after the burst. The data were analyzed with the Common Astronomy Software Applications (CASA) using 3C48 or 3C286 as a flux calibrator (depending on the LST start time of each observation) and J1810+5649 as a gain calibrator. The flux densities and associated uncertainties were determined using the \texttt{imtool} program within the \texttt{pwkit} package\footnote{Available at \url{https://github.com/pkgw/pwkit}.} (version 0.8.4.99; \citealt{pwkit}) and are reported in Table \ref{tab:rad}. The radio light curves are shown in Figure \ref{fig:lc} and the radio spectral energy distributions (SEDs) at the various epochs are shown in Figure \ref{fig:sed1}.

\begin{center}
\setlength\LTcapwidth{2.5in}
\begin{longtable}{lccc}
\caption{Radio Observations}
\label{tab:rad} \\
\hline
\hline\noalign{\smallskip}
$t$ & Frequency & Flux Density \\
(d)         & (GHz)  & ($\mu$Jy)  \smallskip \\
\hline\noalign{}
1.37 & 5.0 & 163 $\pm$ 34 \\ 
1.37 & 7.1 & 232 $\pm$ 22 \\ 
1.35 & 8.5 & 288 $\pm$ 23 \\ 
1.35 & 11.0 & 507 $\pm$ 35 \\ 
\hline\noalign{}
2.50 & 5.0 & 932 $\pm$ 24 \\ 
2.50 & 7.1 & 1310 $\pm$ 20 \\ 
2.49 & 8.5 & 1135 $\pm$ 28 \\ 
2.49 & 11.0 & 946 $\pm$ 25 \\ 
2.47 & 13.5 & 646 $\pm$ 25 \\ 
2.47 & 16.0 & 650 $\pm$ 19 \\ 
2.45 & 19.2 & 553 $\pm$ 34 \\ 
2.45 & 24.5 & 530 $\pm$ 47 \\ 
\hline\noalign{}
6.31 & 1.45 & 300 $\pm$ 90 \\ 
6.31 & 1.77 & 200 $\pm$ 90 \\ 
6.30 & 2.68 & 164 $\pm$ 33 \\ 
6.30 & 3.52 & 165 $\pm$ 43 \\ 
6.29 & 5.0 & 117 $\pm$ 21 \\ 
6.29 & 7.1 & 180 $\pm$ 24 \\ 
6.28 & 8.5 & 262 $\pm$ 41 \\ 
6.28 & 11.0 & 209 $\pm$ 32 \\ 
7.32 & 13.5 & 270 $\pm$ 18 \\ 
7.32 & 16.0 & 237 $\pm$ 23 \\ 
7.30 & 19.2 & 119 $\pm$ 40 \\ 
7.30 & 24.5 & 80 $\pm$ 27 \\ 
\hline\noalign{}
12.50 & 1.45 & 297 $\pm$ 74 \\ 
12.50 & 1.77 & 307 $\pm$ 50 \\ 
12.49 & 2.68 & 621 $\pm$ 31 \\ 
12.49 & 3.52 & 475 $\pm$ 40 \\ 
12.48 & 5.0 & 219 $\pm$ 21 \\ 
12.48 & 7.1 & 185 $\pm$ 21 \\ 
12.47 & 8.5 & 176 $\pm$ 23 \\ 
12.47 & 11.0 & 193 $\pm$ 21 \\ 
12.45 & 13.5 & 176 $\pm$ 23 \\ 
12.45 & 16.0 & 202 $\pm$ 21 \\ 
12.43 & 19.2 & 218 $\pm$ 26 \\ 
12.43 & 24.5 & 147 $\pm$ 38 \\ 
\hline\noalign{}
22.52 & 1.45 & 265 $\pm$ 75 \\ 
22.52 & 1.77 & 346 $\pm$ 62 \\ 
22.51 & 2.68 & 512 $\pm$ 57 \\ 
22.51 & 3.52 & 300 $\pm$ 27 \\ 
22.50 & 5.0 & 229 $\pm$ 31 \\ 
22.50 & 7.1 & 201 $\pm$ 25 \\ 
22.49 & 8.5 & 183 $\pm$ 24 \\ 
22.49 & 11.0 & 132 $\pm$ 30 \\ 
22.47 & 13.5 & 134 $\pm$ 22 \\ 
22.47 & 16.0 & 128 $\pm$ 28 \\ 
22.45 & 19.2 & 159 $\pm$ 38 \\ 
22.45 & 24.5 & 85 $\pm$ 30 \\ 
\hline\noalign{}
48.38 & 1.45 & 142 $\pm$ 47 \\ 
48.38 & 1.77 & 120 $\pm$ 61 \\ 
48.37 & 2.68 & 109 $\pm$ 35 \\ 
48.37 & 3.52 & 72 $\pm$ 24 \\ 
48.36 & 5.0 & 96 $\pm$ 31 \\ 
48.36 & 7.1 & 101 $\pm$ 21 \\ 
48.35 & 8.5 & 84 $\pm$ 25 \\ 
48.35 & 11.0 & 95 $\pm$ 23 \\ 
48.33 & 13.5 & 78 $\pm$ 16 \\ 
48.33 & 16.0 & 97 $\pm$ 21 \\ 
48.31 & 19.2 & 81 $\pm$ 33 \\ 
48.31 & 24.5 & 82 $\pm$ 27 \\ 
\hline\noalign{\smallskip}
\caption[]{VLA observations of GRB 160625B.  All values of $t$ are relative to the LAT trigger time, 2016 June 25 22:43:24.82 UT.}
\end{longtable}
\end{center}

\section{Basic Considerations}\label{sec:bc}

We interpret our multi-wavelength observations using a standard synchrotron emission model \citep{spn98,gs02}. In this model, the emitting electrons are assumed to have been accelerated into a non-thermal distribution $N(\gamma) \propto \gamma^{-p}$ for $\gamma > \gamma_{\text m}$, where $\gamma_m$ is the minimum Lorentz factor of the distribution. The resulting SED is described by three break frequencies (the self-absorption frequency, $\nu_{\text{a}}$, the characteristic synchrotron frequency, $\nu_{\text{m}}$, and the cooling frequency, $\nu_{\text{c}}$) and an overall flux normalization. The temporal evolution of these quantities depends on the circumburst density profile and the outflow geometry. In this section, we estimate basic properties of the afterglow and consider two possible models for the circumburst medium: a constant density ISM profile \citep{spn98} and a wind profile where the density scales as $r^{-2}$ \citep{cl00}.

\subsection{Time of jet break}\label{sec:tjet}
The X-ray, $r^{\prime}$, and $i^{\prime}$ band light curves all steepen at $t\approx25$ d, suggestive of a jet break. The best constraints on the break timing and post-break decline rate come from the $r^{\prime}$ band light curve, which can be fit by two power law segments with a break at $t_{\text{jet}}=27\pm2$ d. Before the break, the decline rate is $\alpha_{\rm1,r}=-0.94\pm0.01$; after the break, it steepens to $\alpha_{\rm2,r}=-2.3\pm0.4$ ($\Delta \alpha_{\rm12,r}=-1.4\pm0.4$). The steep post-break decline rate and the lack of flattening at late times indicate that the GRB host contributes negligibly to the total flux. By $t=t_{\rm jet}$, we expect $\nu_{\rm m}$ to be located below the optical band, and the $r^{\prime}$ band light curve should therefore evolve as $t^{-p}$ after the jet break \citep{sari99}. We therefore estimate $p\approx2.3$ for the non-thermal electron distribution.

The radio observations also show evidence of a jet break, as the flux declines at all frequencies between 22 d and 48 d. The higher frequencies ($\nu > 7$ GHz) prefer a significantly earlier jet break time than the optical and X-ray observations, $t_{\text{jet}}\approx12$ d; other effects dominate the emission at frequencies below 7 GHz during this time range (see Section \ref{sec:radio}.) Such an earlier jet break would require the presence of an additional component to explain the smooth decline of the optical and X-ray emission at $t\approx12-27$ d. However, this explanation is disfavored due to its increased complexity and as there are other signs of unusual variability in the radio, we take $t_{\text{jet}}\approx25$ d as preferred by the optical and X-ray data. 

\subsection{Circumburst density profile, location of $\nu_{\rm c}$, host extinction}\label{sec:nuc}
Prior to $t=t_{\text{jet}}$, the optical and X-ray light curves can each be fit with a single power law. The $i^{\prime}$ band light curve has a similar decline rates to the $r^{\prime}$ band light curve, $\alpha_{1,\text{i}}=-0.94 \pm 0.02$, while the X-ray light curve declines more steeply, with $\alpha_{1,\text{XRT}}=-1.24 \pm 0.02$. A natural explanation for this in the context of the synchrotron model is that the cooling break ($\nu_{\text c}$) is located between the optical and X-ray bands. The predicted decline rate for $\nu<\nu_{\text c}$ depends on the circumburst density profile and is $\alpha_{\text{ISM}}=3(1-p)/4$ for an ISM profile and $\alpha_{\text{wind}}=(1-3p)/4$ for a wind profile \citep{gs02}. Using the $r^{\prime}$ band light curve, we find $p=2.25 \pm 0.02$ for the ISM case and $p=1.59 \pm 0.02$ for the wind case. For both profiles, the predicted decline rate for $\nu>\nu_{\text c}$ is $\alpha=(2-3p)/4$ and the X-ray decline rate implies $p = 2.32 \pm 0.03$. The pre-jet break optical and X-ray observations are thus only self-consistent if the circumburst medium is ISM-like rather than wind-like, giving $p\approx2.3$ in agreement with the value derived from the post-jet break decline rate in Section \ref{sec:tjet}. We therefore only consider the ISM profile for our detailed modeling in Sections \ref{sec:mod} and \ref{sec:radio}. 

We can also use the inferred value of $p$ and the optical/NIR spectral energy distribution to constrain the amount of extinction in the GRB host. For $\nu < \nu_c$ and zero extinction, the predicted spectral index is $\beta = -0.65$ for $p=2.3$. Fitting the RATIR $rizYJH$ data points at 1.468 d \citep{wat16}, we find a spectral index of $\beta_{\text{NIR}}=-0.68\pm0.07$, consistent with this value. We see a slightly steeper $r-g$ spectral index in MITSuME observations at 0.731 d \citep{kur16}, $\beta_{rg}=-1.0\pm0.2$. This indicates a small total amount of extinction along the line of sight, consistent with the expected amount of Galactic extinction (Section \ref{sec:opt}) and little to no extinction in the GRB host galaxy. The spectral index in the XRT 0.3$-$10 keV band is $\beta_{\rm X}=-0.86_{-0.10}^{+0.09}$, which is intermediate between the values expected  for $p\approx2.3$ when $\nu_{\rm X} < \nu_{\rm c}$ ($\beta_{\rm X}\approx-0.65$) and $\nu_{\rm X} > \nu_{\rm c}$ ($\beta_{\rm X}\approx-1.15$). This may indicate that $\nu_{\rm c}$ is located only slightly below the X-ray band, as the spectrum is expected to transition smoothly from one power law index to the other around each break frequency. The NIR to X-ray spectral index is $\beta_{NIR-X}=-0.71\pm0.01$, slightly steeper than expected if $\nu_{\rm X} < \nu_{\rm c}$ for $p\approx2.3$. Therefore $\beta_{NIR-X}$ is also consistent with $\nu_c$ being located just below the X-ray band.

\subsection{Multiple radio components}\label{sec:mrc}
\begin{figure} 
\centerline{\includegraphics[width=3.8in]{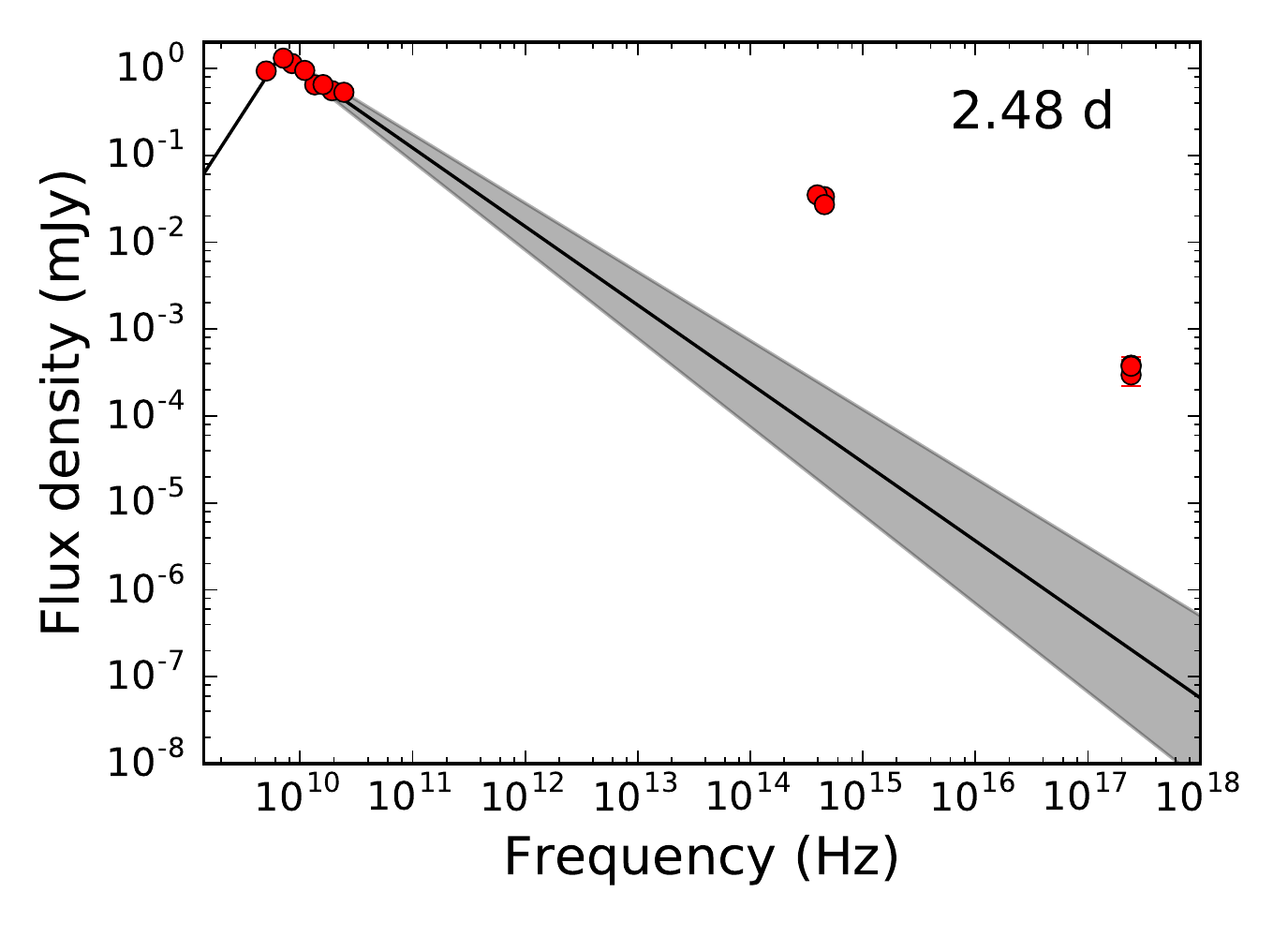}} 
\vspace{-0.15in}
\centerline{\includegraphics[width=3.7in]{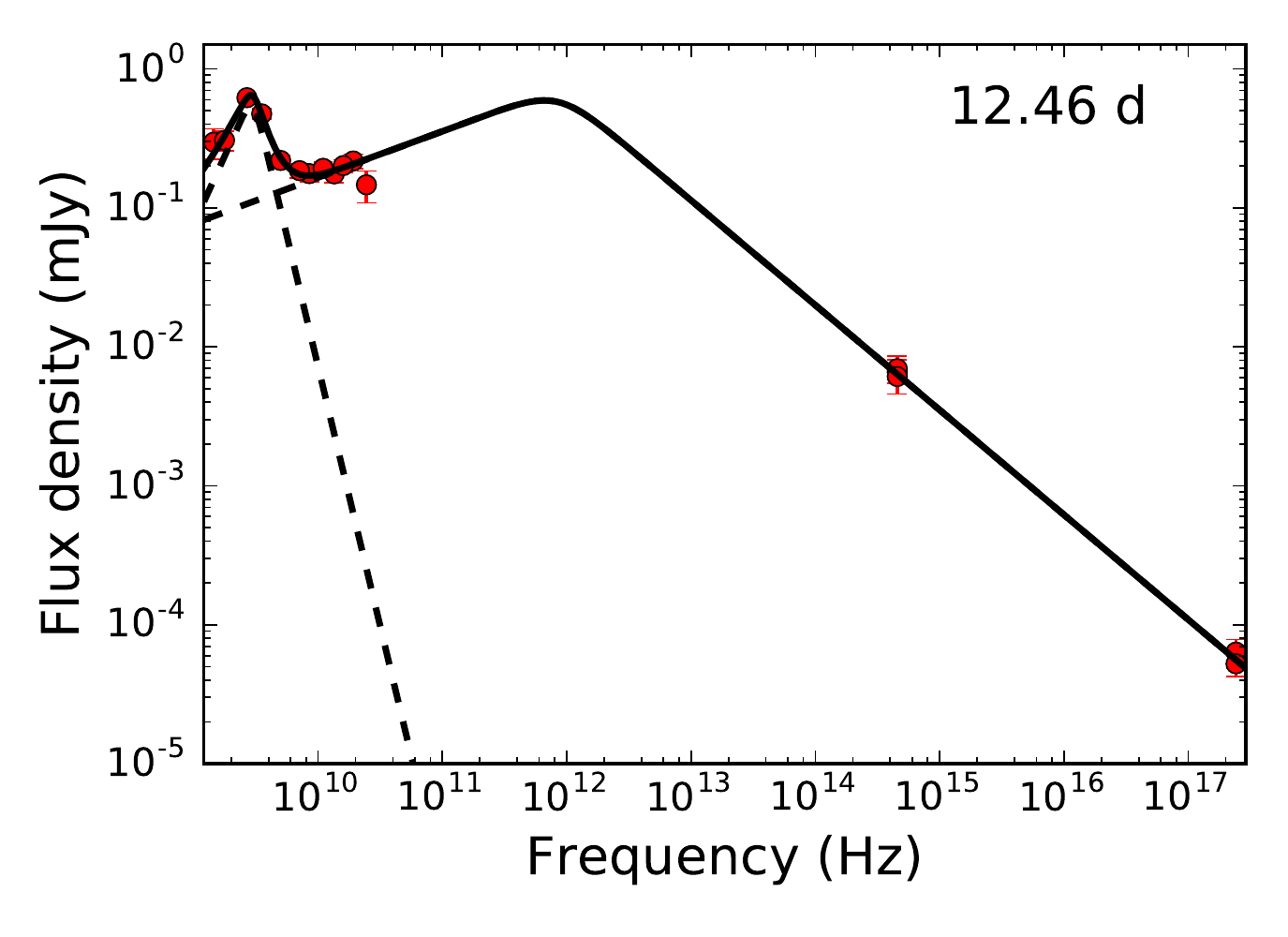}}
\vspace{-0.15in}
\caption{Top: The radio to X-ray spectral energy distribution at 2.48 d. The radio data are fit with a broken power law and the spectral index above 6 GHz is extended to the optical and X-ray bands (black line, shaded region indicates the 1$\sigma$ uncertainty in the fit). The fit underpredicts the optical and X-ray emission by several orders of magnitude, indicating that the radio emission is dominated by a separate component at this time. Bottom: The radio to X-ray spectral energy distribution at 12.46 d fit with two components. The radio data above 8 GHz connect simply to the optical and X-ray data with a $\nu^{1/3}$ power law transitioning to a $\nu^{-0.75}$ power law, as expected for the forward shock. The radio data below 8 GHz require a second, extremely spectrally narrow component that does not connect simply to the FS or to the component dominating the radio emission at 2.48 d.}
\label{fig:rxrt}
\end{figure}

The radio emission at $t=2.48$ d is dominated by a single component with a spectral peak around 6 GHz. If the emission is fit with a broken power law and the spectral index above the peak is extrapolated to high frequencies, this component underpredicts the observed optical and X-ray emission by several orders of magnitude (Figure \ref{fig:rxrt}; top). We therefore conclude that a separate mechanism is required to explain the radio emission at $t\leq7$ d and show in Section \ref{sec:rs} that this component is consistent with a reverse shock. The peak of this component must be above 11 GHz at 1.36 d, implying that the peak frequency evolves faster than $t^{-1}$. This means that $\nu_p \lesssim 2$ GHz at 6.8 d and $\nu_p\lesssim1$ GHz at 12.46 d, indicating that this component cannot contribute significantly to the observed radio emission after 7 d.

We also observe a low-frequency rebrightening at $12-22$ d peaked at $\sim3$ GHz, which appears distinct from higher-frequency emission at that time (Figure \ref{fig:rxrt}; bottom). The high-frequency emission is broadly consistent with expectations for the FS. The low-frequency emission cannot be the same component dominating the radio emission before 7 d unless that component's peak frequency were to start increasing in time after 7 d; such behavior is not predicted for either FS or RS emission and would be unprecedented in GRB afterglow studies. This component is also too spectrally narrow for standard synchrotron emission: for the broken power law fit in Figure \ref{fig:rxrt} we find that the spectral index is $\beta_1=3.0 \pm 0.1$ below the peak and $\beta_2=-3.7\pm0.6$ above it. Together, these properties suggest distortion of the intrinsic low-frequency radio SED by interstellar scintillation (ISS) as the emission propagates through the turbulent Galactic ISM (see review by \citealt{rick90}). ISS is known to cause strong, uncorrelated flux density variations in GRB afterglows and other sufficiently compact radio sources and should be carefully considered before claiming that observed rapid spectral and temporal variations require exotic new effects intrinsic to the GRB. We discuss ISS and other possible origins of this component in more detail in Section \ref{sec:rad}.

\section{Forward Shock Model}\label{sec:mod}

\begin{figure*} 
\centerline{\includegraphics[width=2.0in]{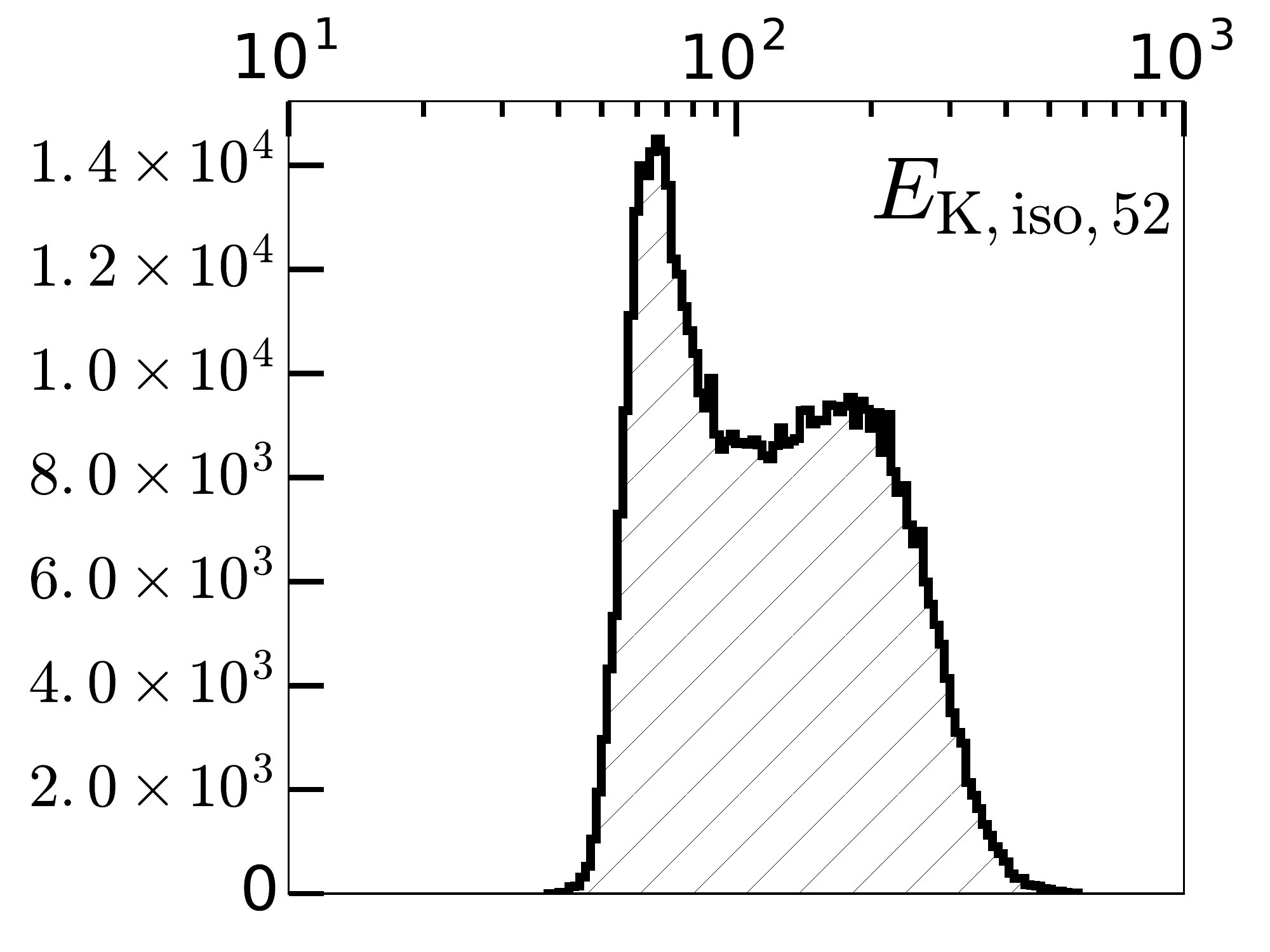}\includegraphics[width=2.0in]{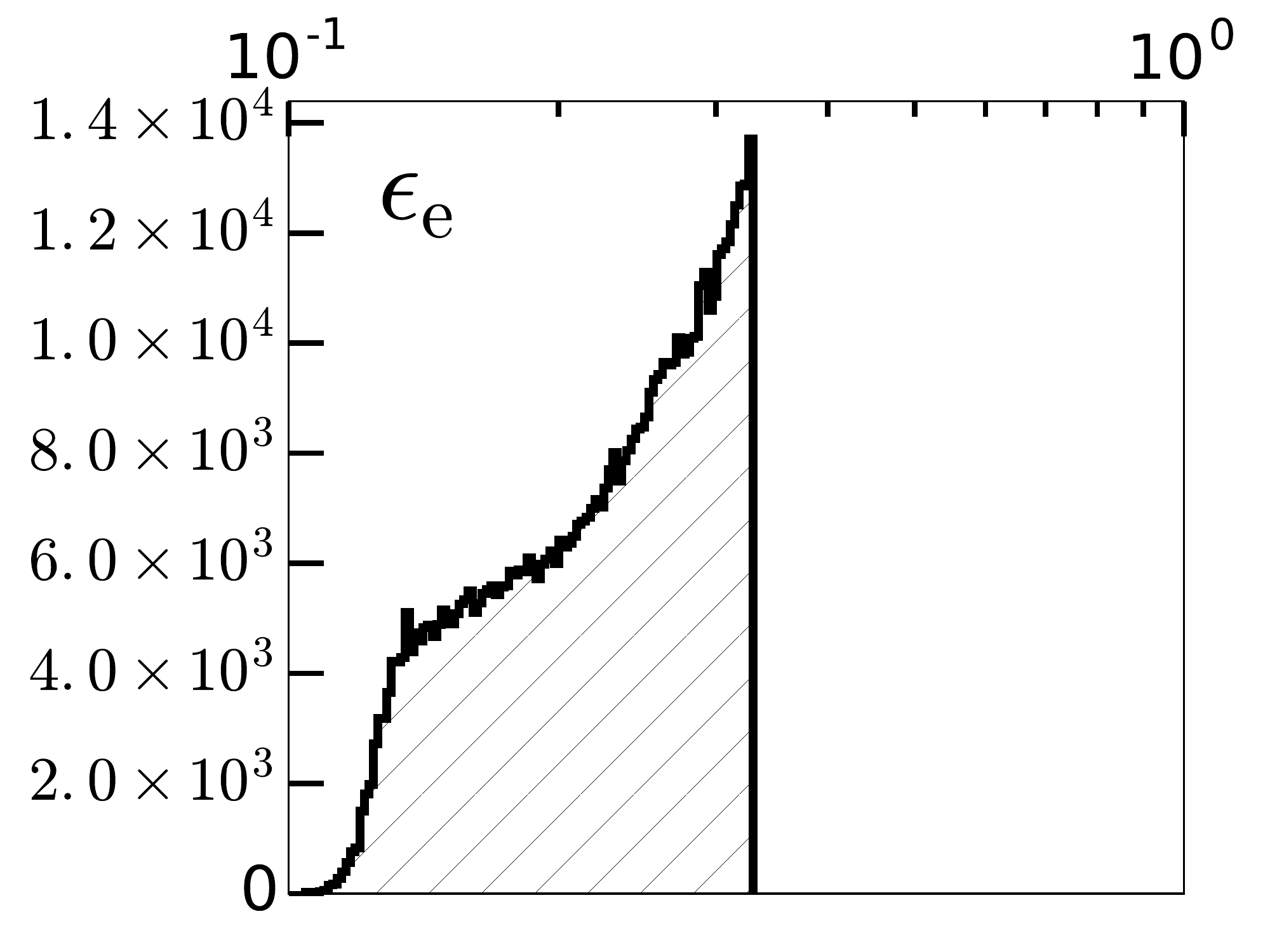}\includegraphics[width=2.0in]{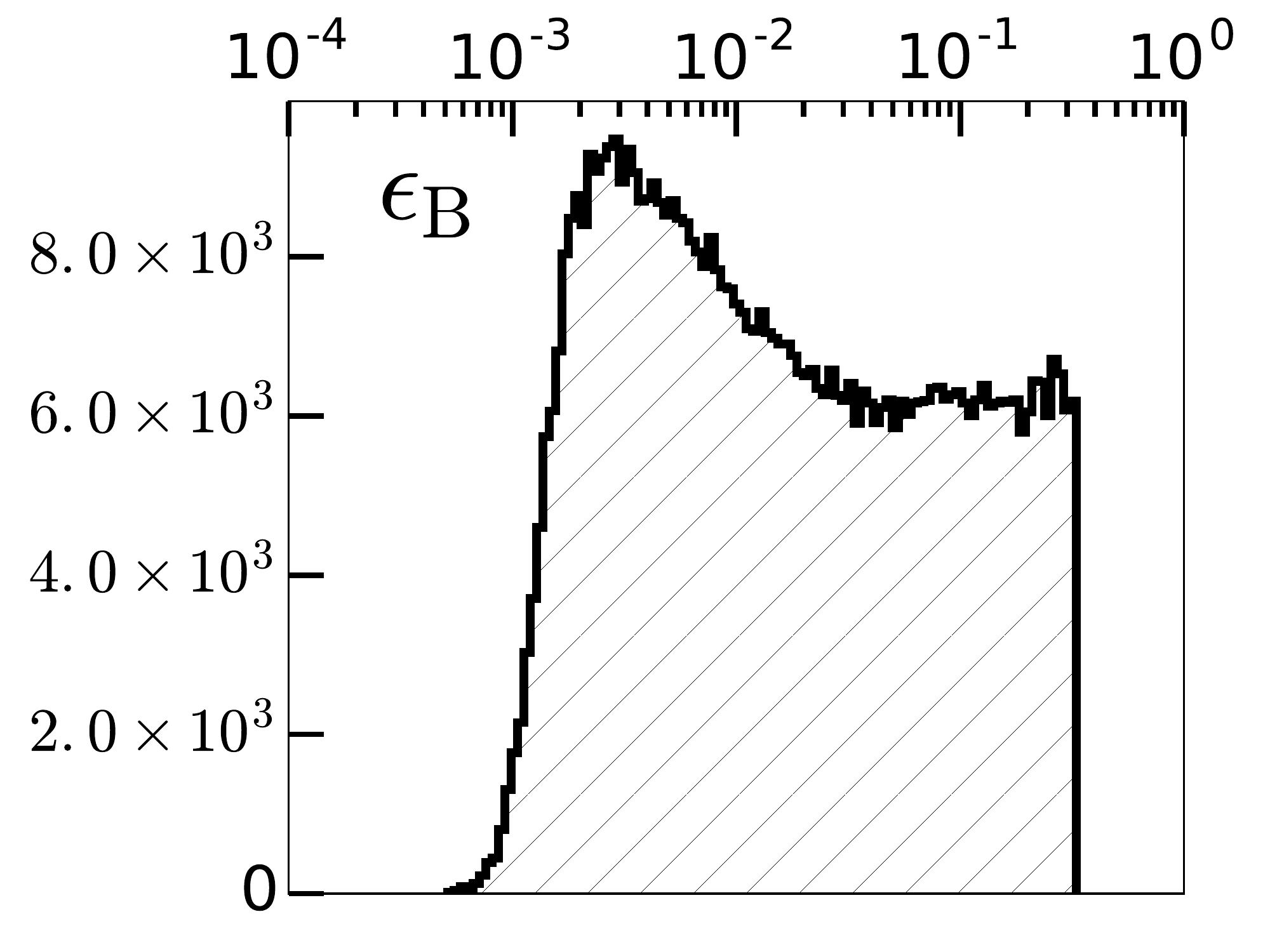}\includegraphics[width=2.0in]{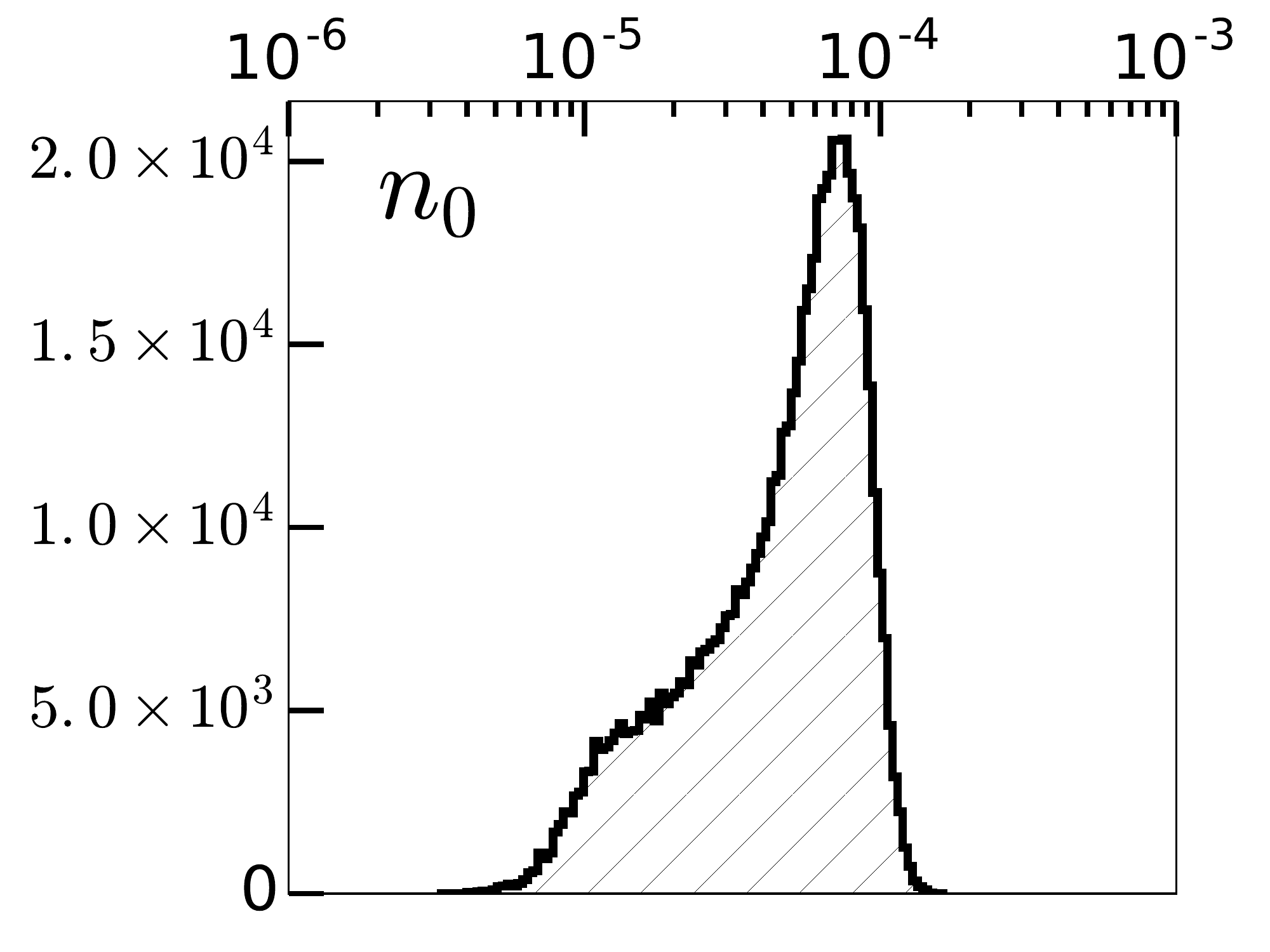}}
\centerline{\includegraphics[width=2.0in]{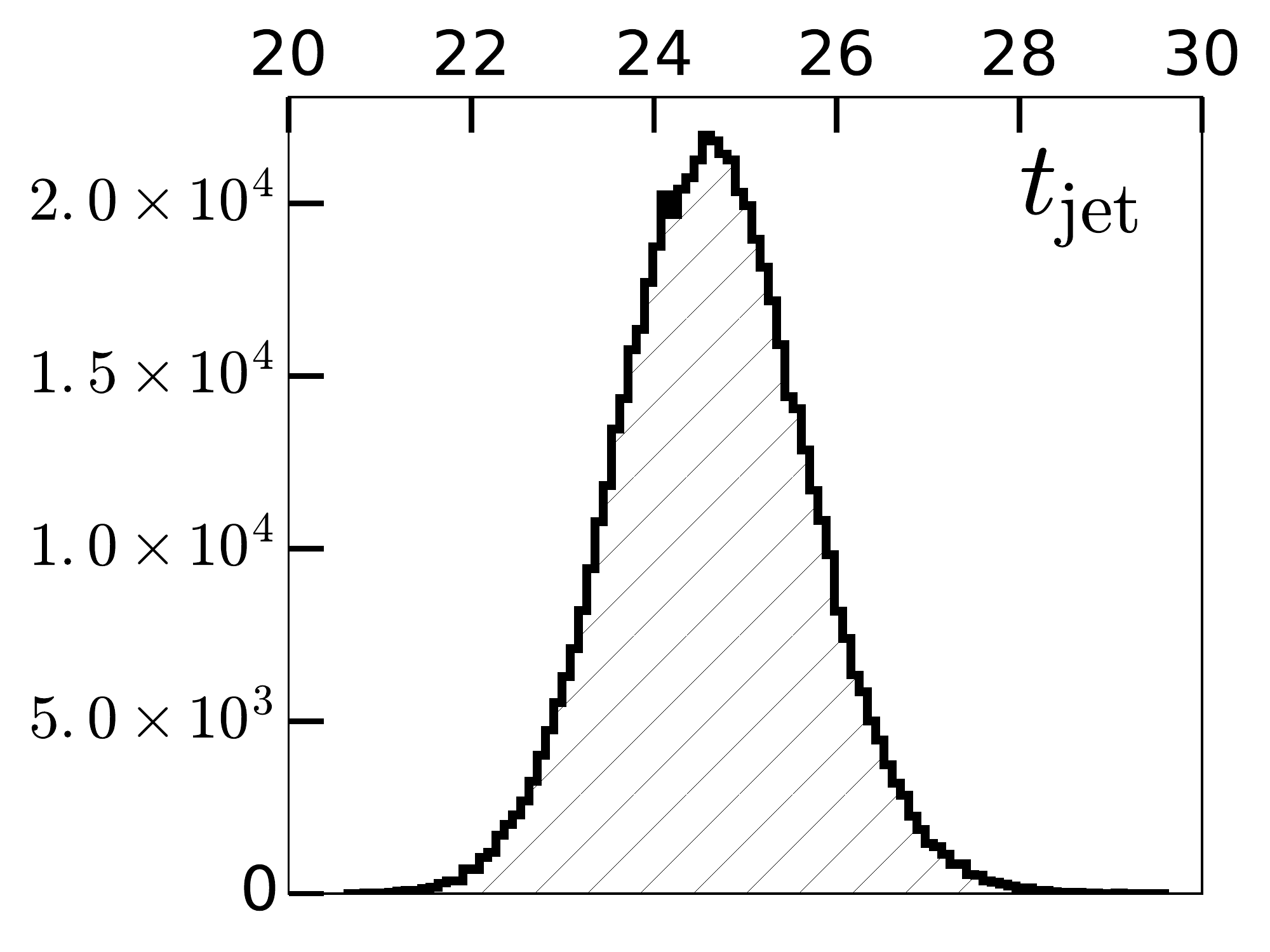}\includegraphics[width=2.0in]{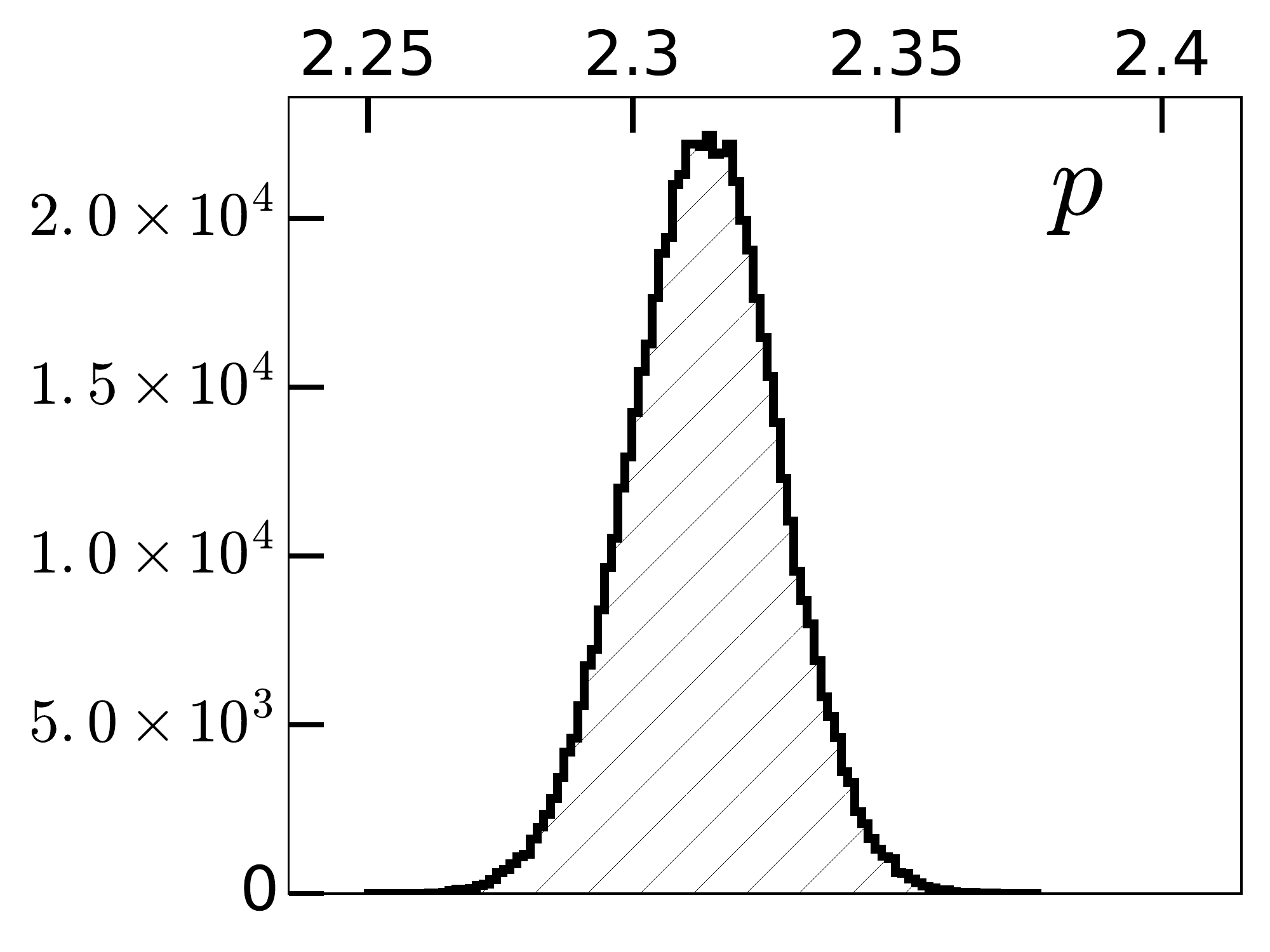}\includegraphics[width=2.0in]{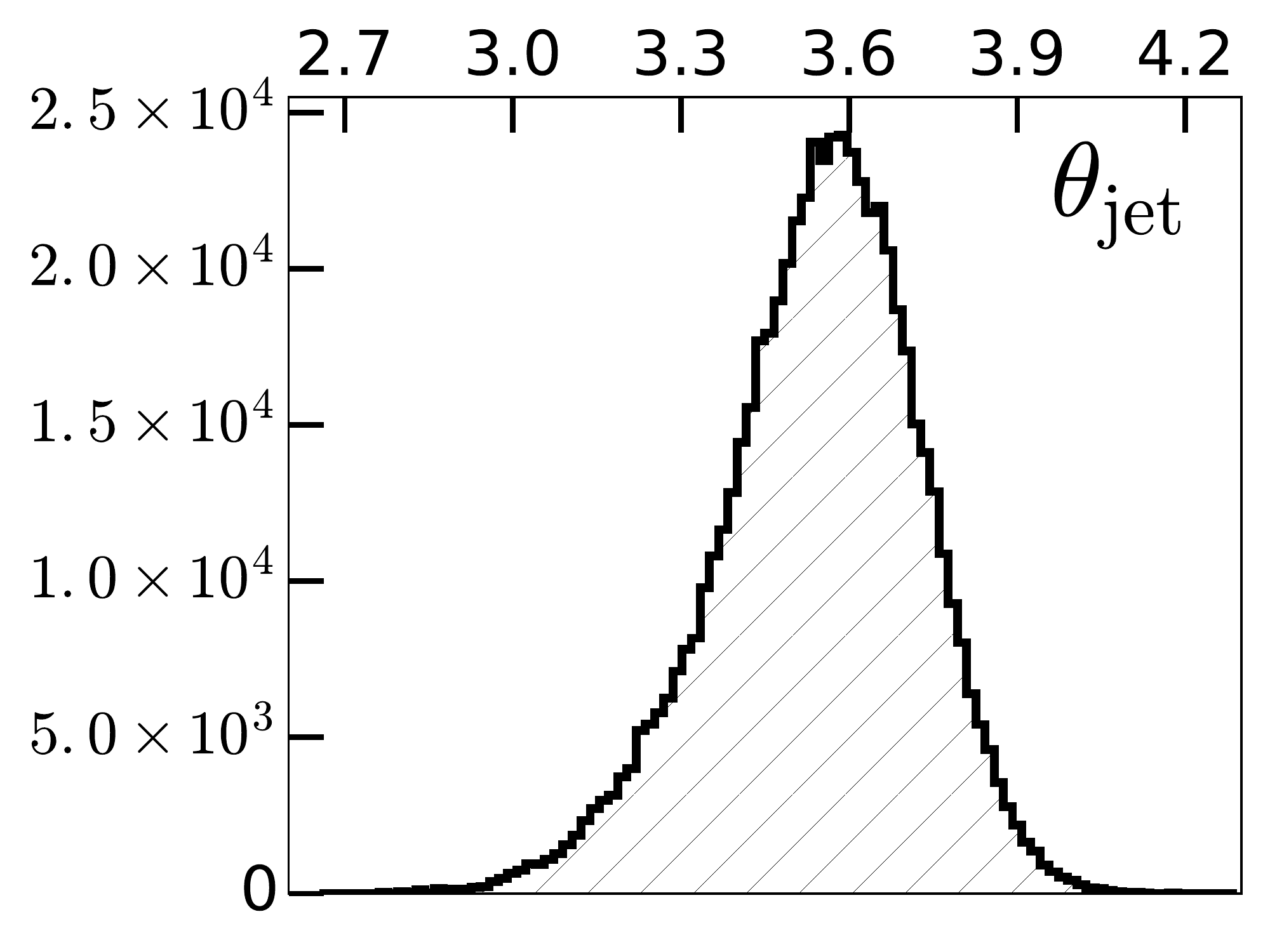}\includegraphics[width=2.0in]{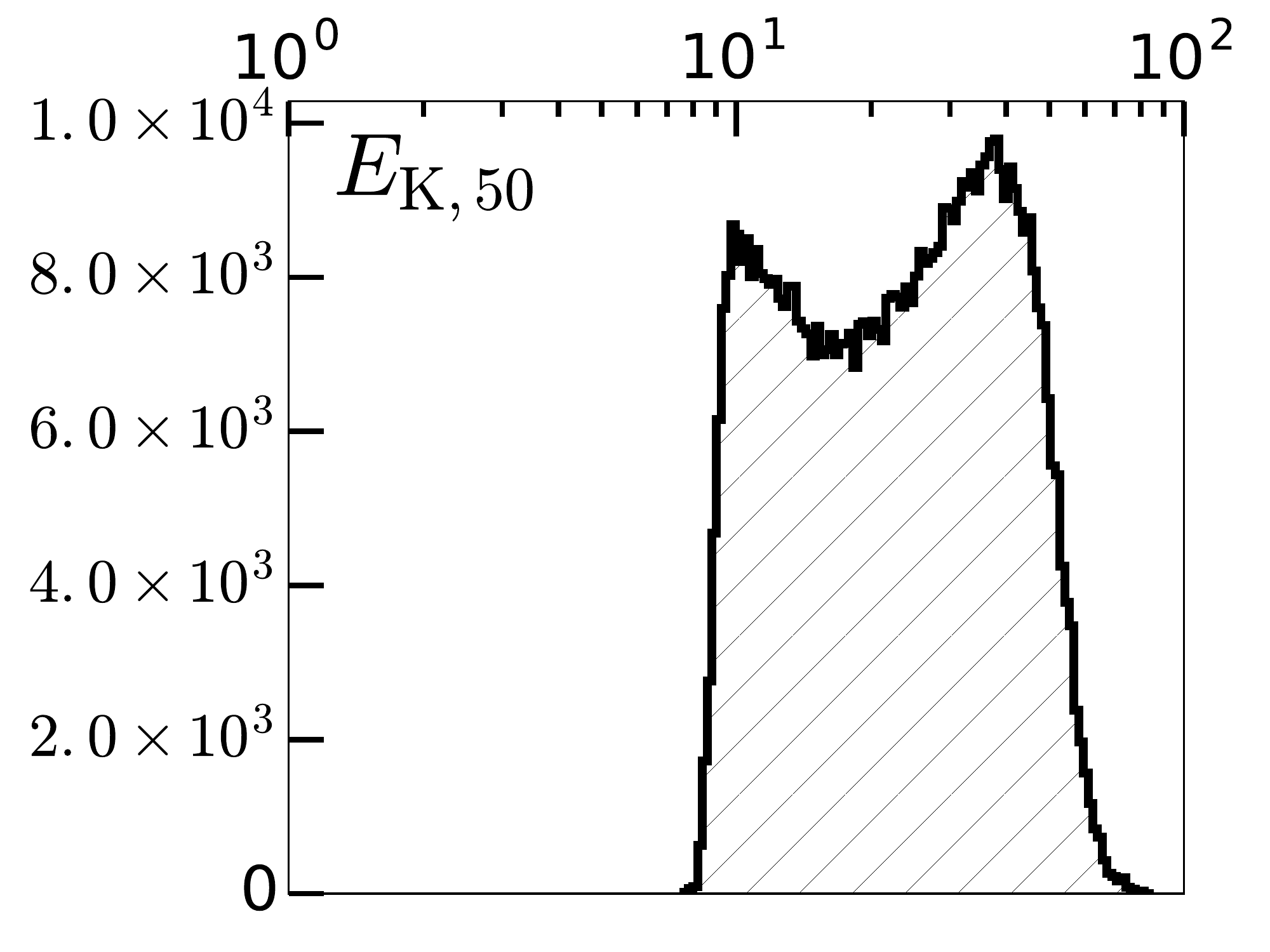}}
\caption{Individual parameter probability density functions for the FS model discussed in Section \ref{sec:mod}. We have followed \cite{lbm+15} in restricting $\epsilon_e < \onethird$ and $\epsilon_B < \onethird$.}
\label{fig:hists}
\end{figure*}

\begin{figure*} 
\centerline{\includegraphics[width=2.3in]{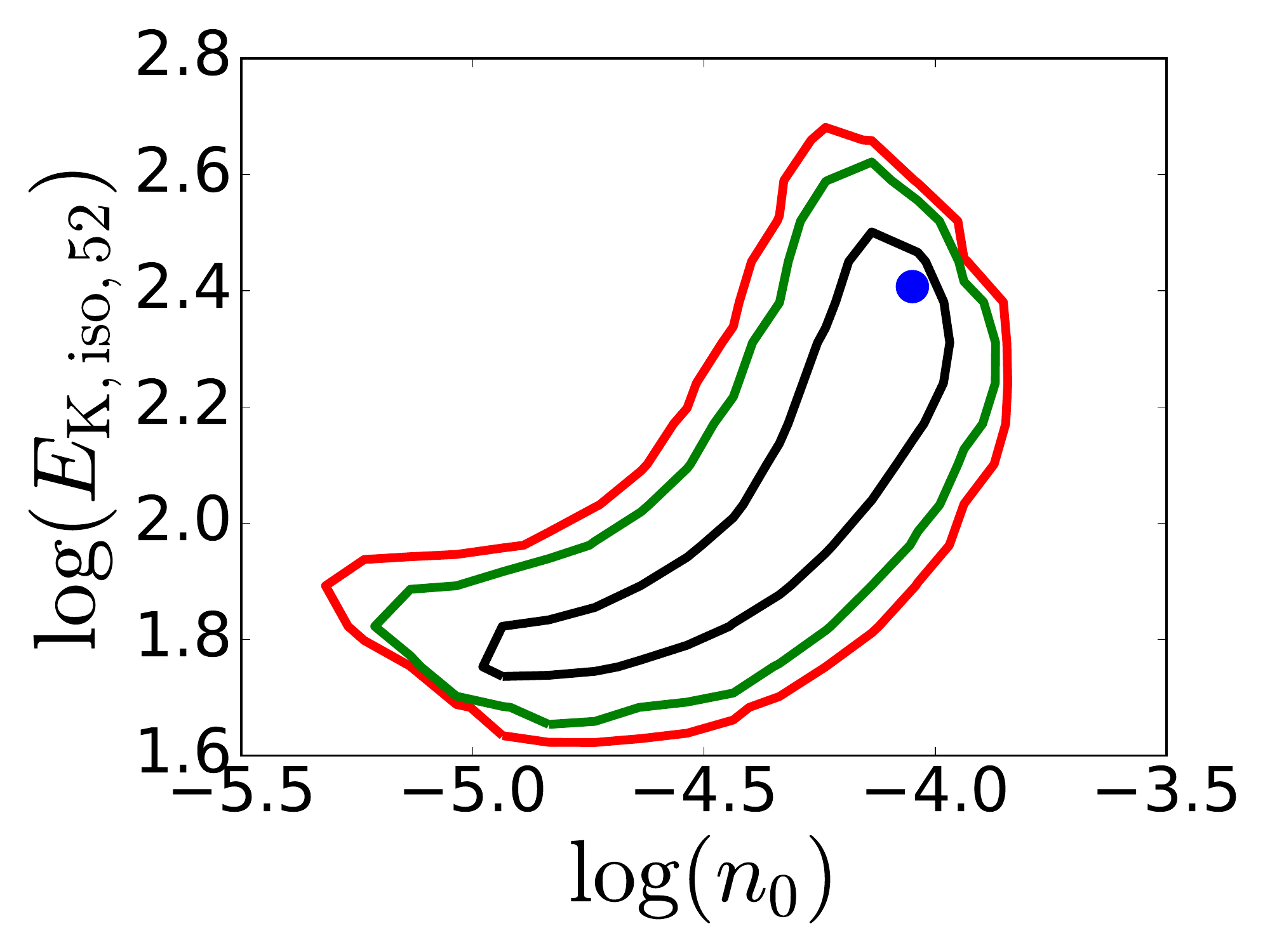}\includegraphics[width=2.3in]{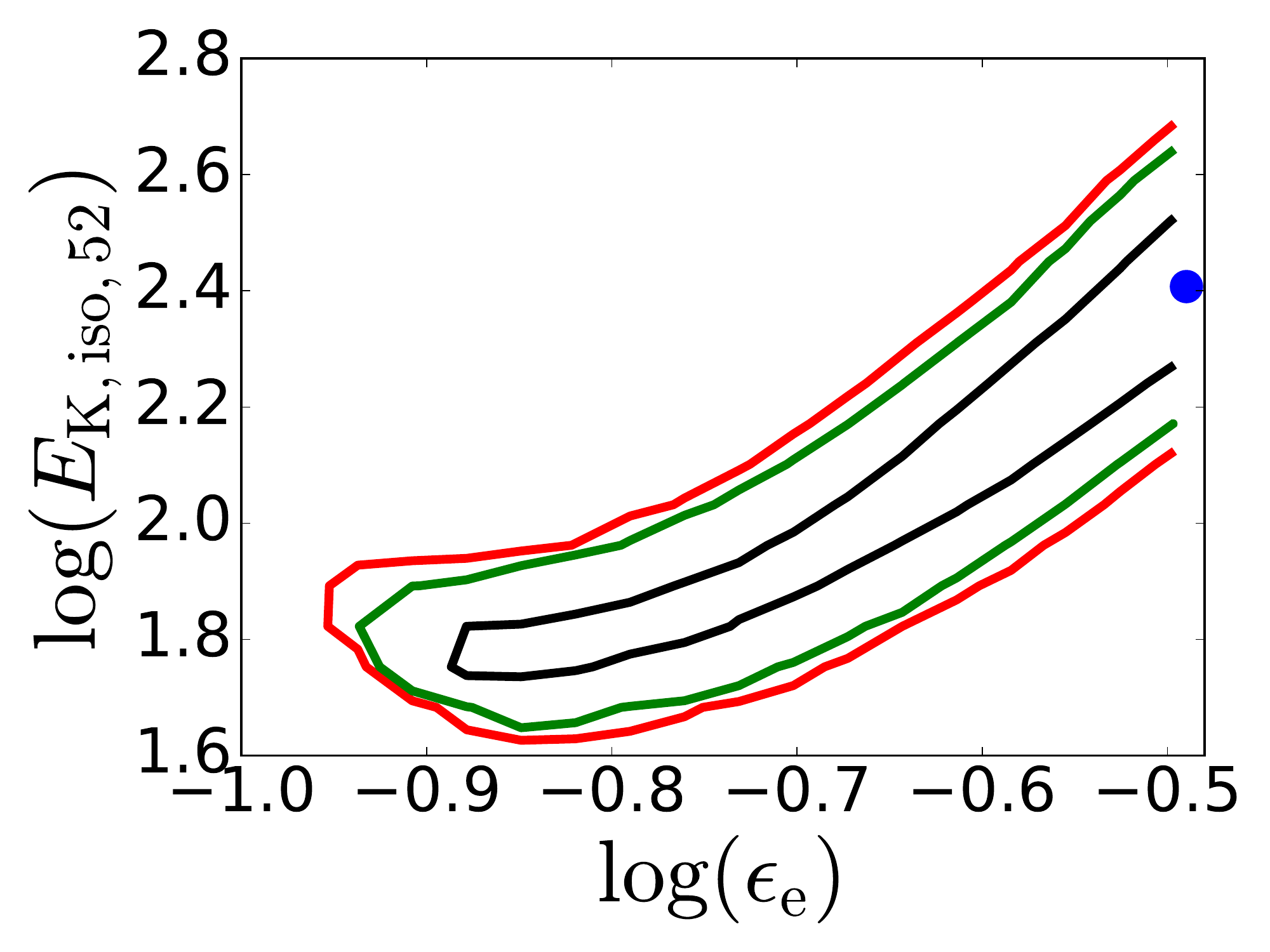}\includegraphics[width=2.3in]{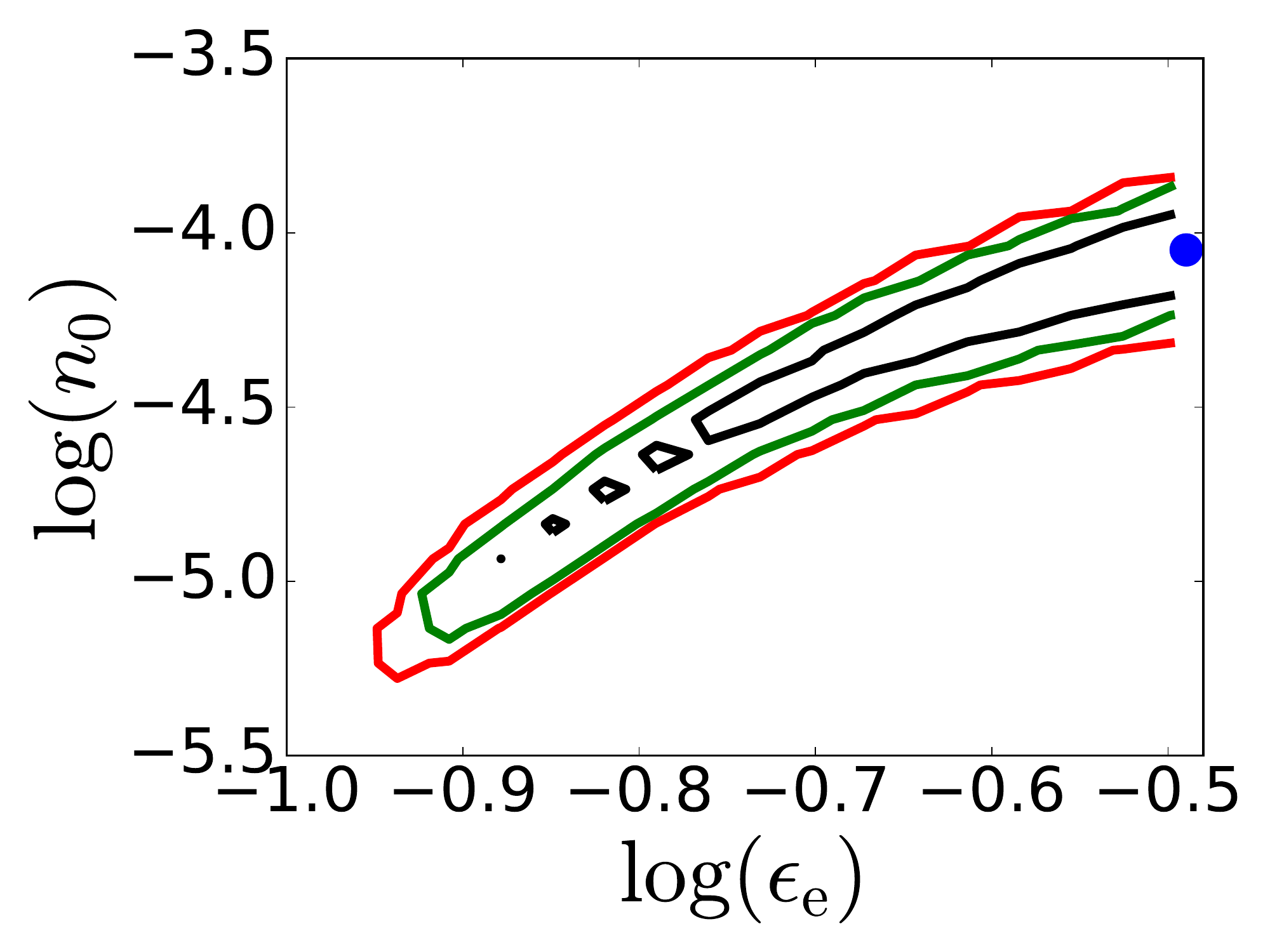}}
\centerline{\includegraphics[width=2.3in]{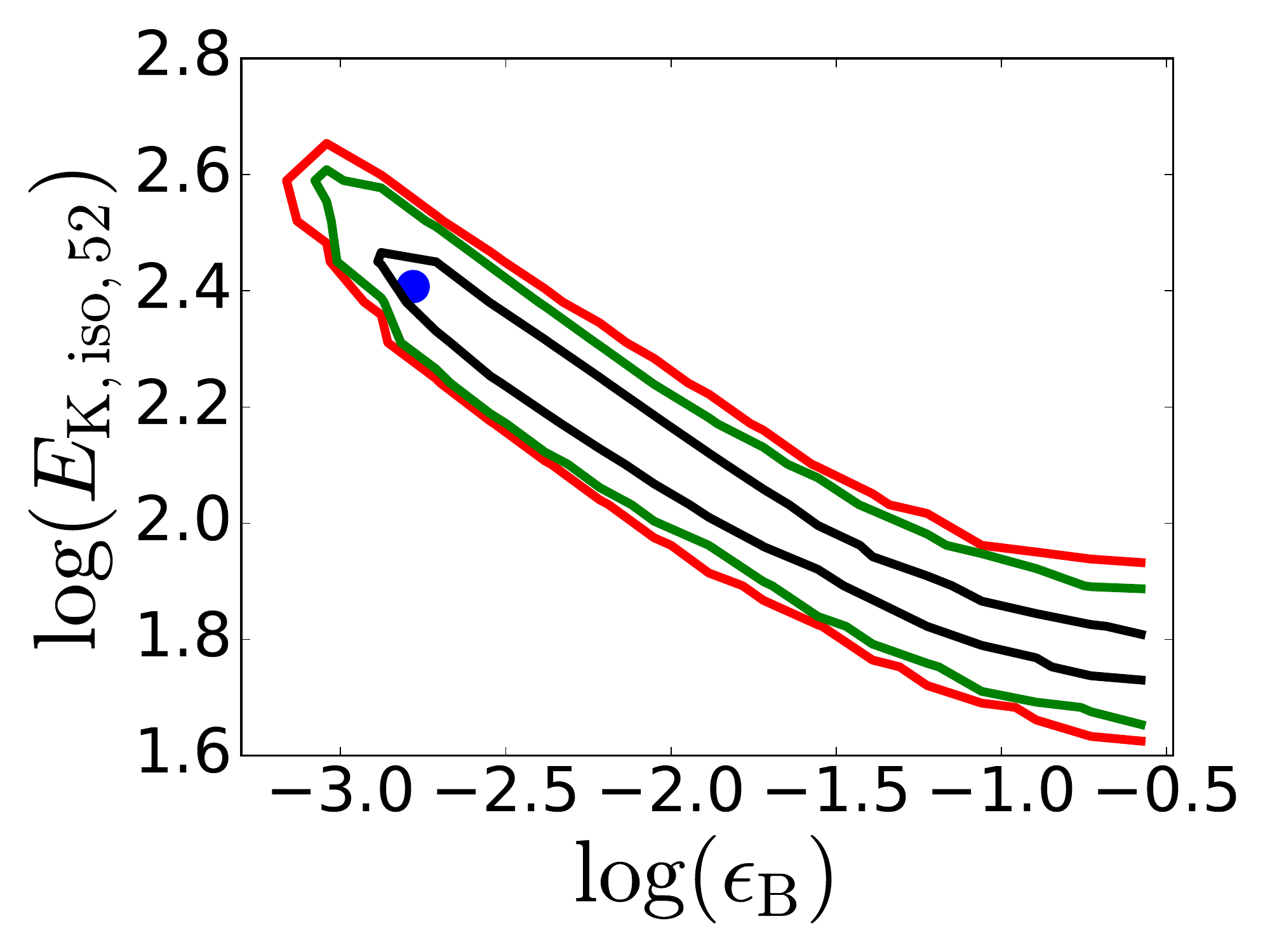}\includegraphics[width=2.3in]{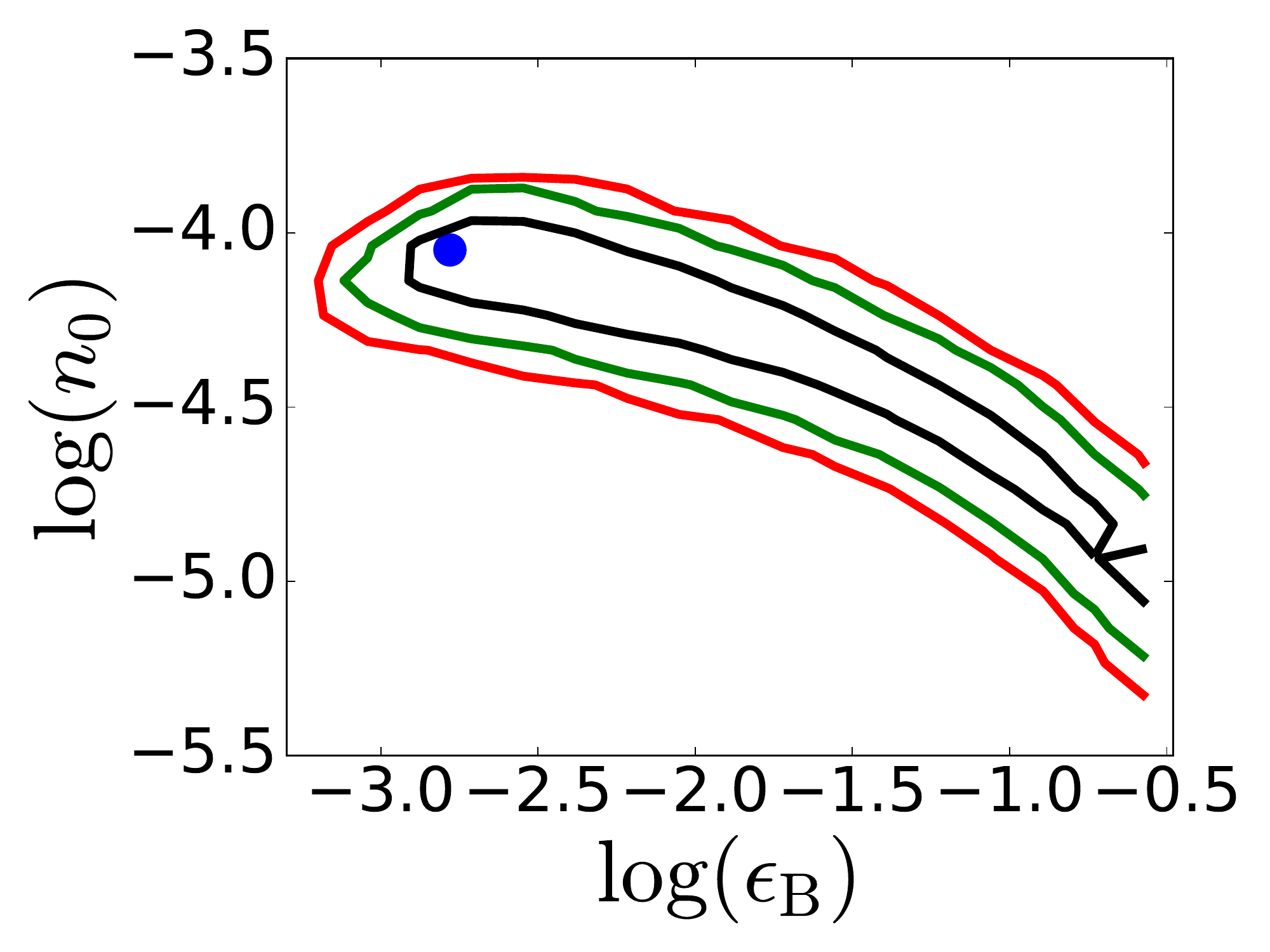}\includegraphics[width=2.3in]{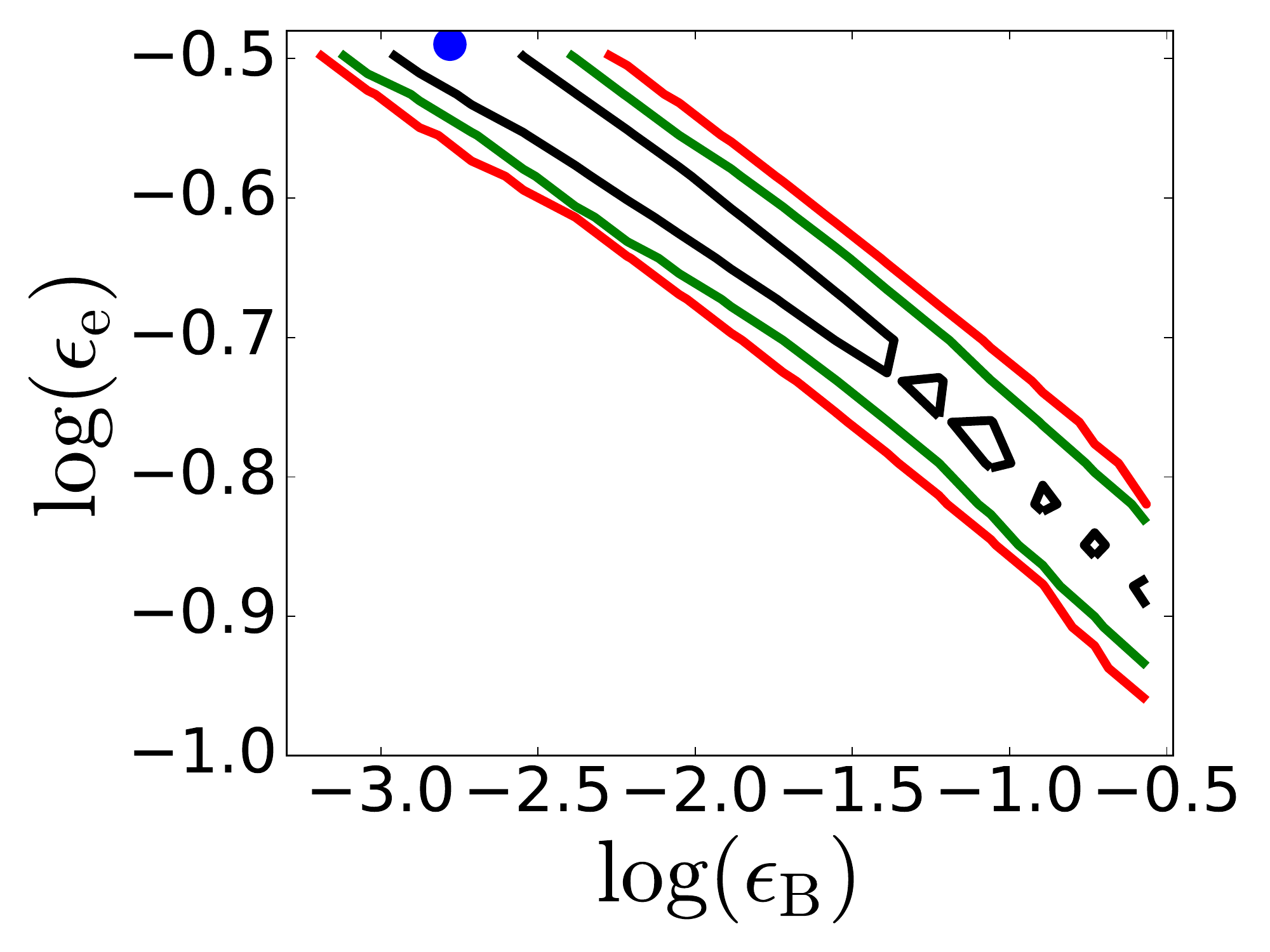}} 
\caption{Physical parameter correlations for the FS model discussed in Section \ref{sec:mod}. The $1\sigma$ (black), $2\sigma$ (green), and $3\sigma$ (red) contours of the parameter distributions are shown, along with the maximum likelihood model (blue points). The degeneracies arise because $\nu_{\rm a}$ of the FS is located below the radio band throughout our observations and is therefore only bounded at the upper end, $\nu_a\lesssim 1$ GHz.}
\label{fig:corrs}
\end{figure*}

Motivated by these basic considerations, we model the afterglow as synchrotron emission resulting from the FS between the jet ejecta and the surrounding medium, including the effects of inverse Compton cooling \citep{se01, gs02}. Our modeling framework is described in detail in \cite{lbt+14} and \cite{lbm+15} and uses the Python package {\tt emcee} \citep{for13} to fully explore parameter space and uncover correlations between physical parameters. The model parameters are the isotropic-equivalent ejecta kinetic energy ($E_{\text{K,iso}}$), the circumburst density ($n_0$), the electron energy index ($p$), the jet break time ($t_{\text{jet}}$), and the fraction of the shock energy imparted to electrons ($\epsilon_e$) and magnetic fields ($\epsilon_B$). We include a correction for Galactic extinction but fix the extinction in the GRB host to $A_V=0$, as the data strongly prefer negligible host extinction if this parameter is allowed to vary freely (consistent with Section \ref{sec:nuc}). We also require $\epsilon_e<\frac{1}{3}$ and $\epsilon_B<\frac{1}{3}$, their equipartition values. This is commonly done to partially break parameter degeneracies that arise when one or more of the FS break frequencies is not well constrained (e.g. \citealt{lbm+15}) and is consistent with recent work that finds most GRBs have $\epsilon_e=0.13-0.15$ \citep{ben17}. We exclude the radio data at early times ($t<12$ d) and all data at frequencies below 7 GHz because other components dominate this emission (Section \ref{sec:radio}). We also exclude the $U$ band data due to the systematic uncertainties discussed in Section \ref{sec:uvot}. The parameters for our best-fit model ($\chi^2=7.56$ for 6 degrees of freedom) are listed in Table \ref{tab:mod} and the model light curves are shown in Figure \ref{fig:lc}. All data points excluded from our model fitting are marked with open symbols in Figure \ref{fig:lc}. The full marginalized posterior probability density functions for each model parameter and two additional derived parameters (the jet opening angle, $\theta_{\rm jet}$, and the beaming-corrected kinetic energy, $E_K$) are given in Figure \ref{fig:hists}. Correlations between the physical parameters $E_{\text{K,iso}}$, $n_0$, $\epsilon_e$, and $\epsilon_B$ are shown in Figure \ref{fig:corrs}.

The self-absorption frequency $\nu_{\rm a}$ is located below the radio band for the entirety of our observations and is therefore poorly constrained. This creates degeneracies between $\epsilon_e$, $\epsilon_B$, $n_0$, and $E_{\rm K,iso}$, as illustrated in Figure \ref{fig:corrs}. This also leads to a large uncertainty in the strength of inverse Compton cooling, with possible Compton $Y$ parameter values ranging from $Y\approx0.2$ (mildly significant cooling) to $Y\approx20$ (strong cooling). Our best-fit model has $Y\approx 3.7$, which is comparable to the value recently found for GRB 160509A ($Y\approx2.4$) and corresponds to moderately significant cooling \citep{lab+16}. We find $p=2.31\pm0.01$ and $t_{\text jet}=25\pm1$ days, in agreement with the arguments presented in Section \ref{sec:bc}. The kinetic energy of the outflow is $E_{\text{K,iso}}=(1.1^{+1.0}_{-0.5}) \times 10^{54}$ erg, similar to the energy released in the prompt emission of this GRB, $E_{\gamma,{\rm iso}}\approx3\times10^{54}$ erg \citep{zzc+16}. This implies a high radiative efficiency for the burst of $\eta_{\gamma} = E_{\gamma, {\rm iso}}/(E_{\rm K, iso}+E_{\gamma, {\rm iso}}) = 0.73^{+0.10}_{-0.14}$, which is within the range of efficiencies found for long GRBs in previous work \citep{zhang07}. The beaming-corrected outflow kinetic energy is $(2.3^{+1.8}_{-1.2}) \times 10^{51}$ erg. The density implied by the model is quite low, $n_0 = (5\pm3) \times10^{-5}$ cm$^{-3}$. Previous studies have found that the circumburst density varies widely among long GRBs, with estimates for individual bursts ranging from $10^{-5}$ to $10^3$ cm$^{-3}$ \citep{lbt+14,lbm+15}. GRB 130427A and GRB 160509A, which both had strong detections of RS emission in the radio, had very low densities of $\approx10^{-3}$ cm$^{-3}$, suggesting that low-density environments may be required to produce observable, long-lasting RS emission \citep{lbz+13,lab+16}. As we will see in Section \ref{sec:rs}, GRB 160625B likely also has a strong RS.

\section{Multiple radio components}\label{sec:radio}

The early radio observations ($t<12$ d) at all frequencies and the low-frequency radio observations ($\nu < 7$ GHz) at all times are not well-fit by the FS model discussed in Section \ref{sec:mod}. A natural explanation for the radio excess at early times is emission from a RS. As a RS alone cannot explain all of the data, we also consider how propagation through the interstellar medium of the Galaxy affects the radio emission via scintillation.

\subsection{Early Radio Emission: A Reverse Shock}\label{sec:rs}

\begin{figure*} 
\centerline{\includegraphics[width=2.5in]{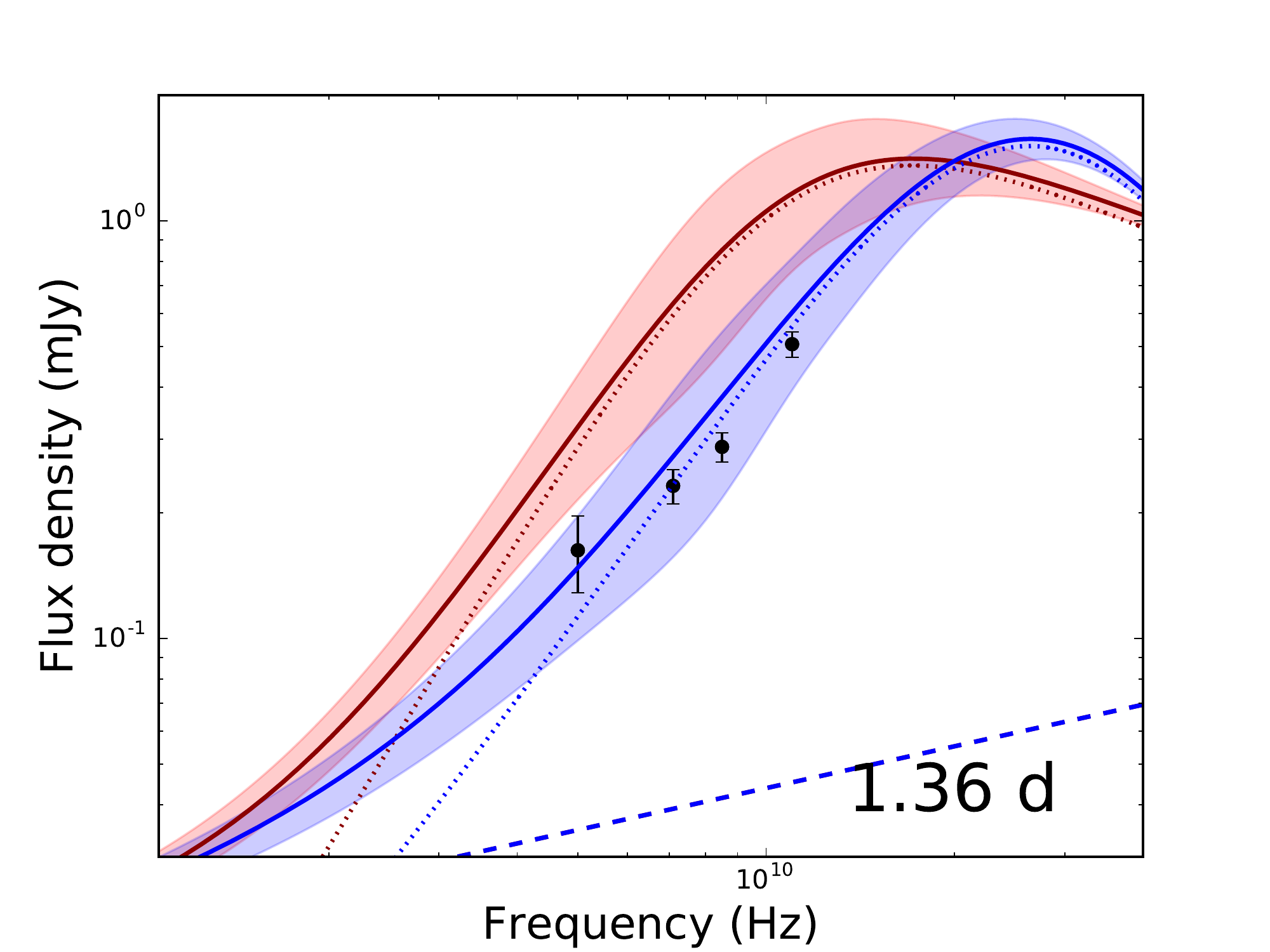}\includegraphics[width=2.5in]{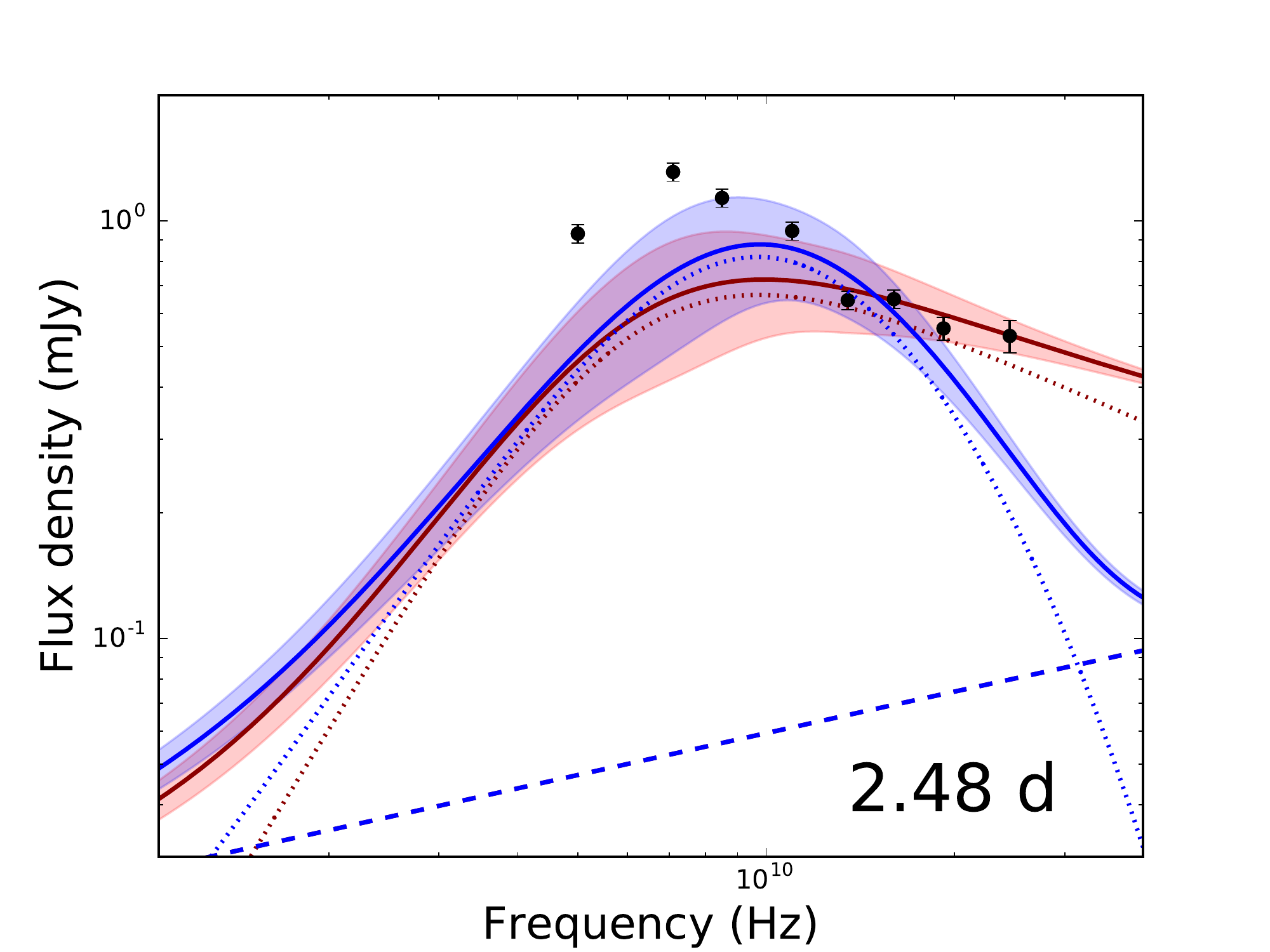}\includegraphics[width=2.5in]{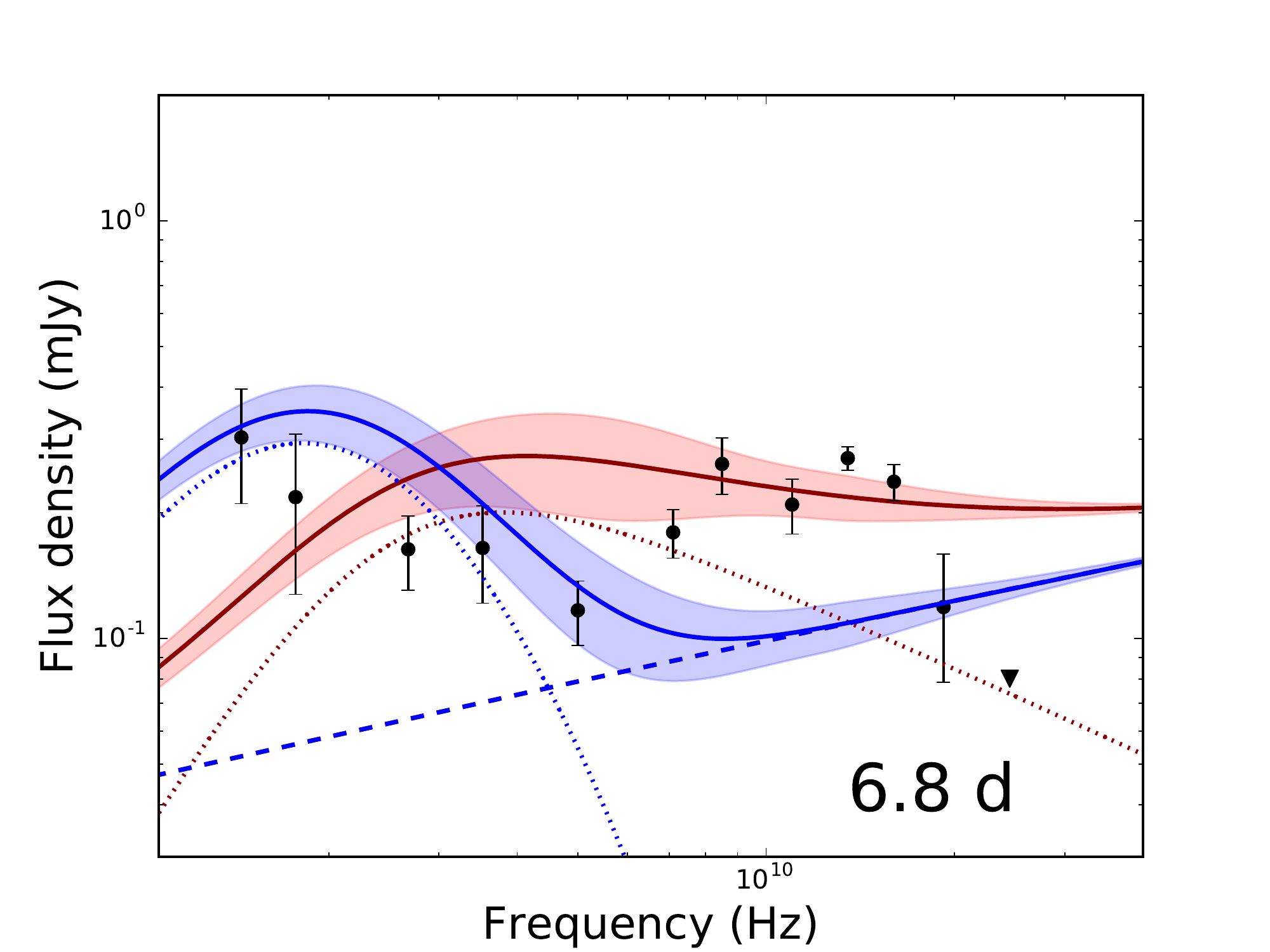}}
\centerline{\includegraphics[width=2.5in]{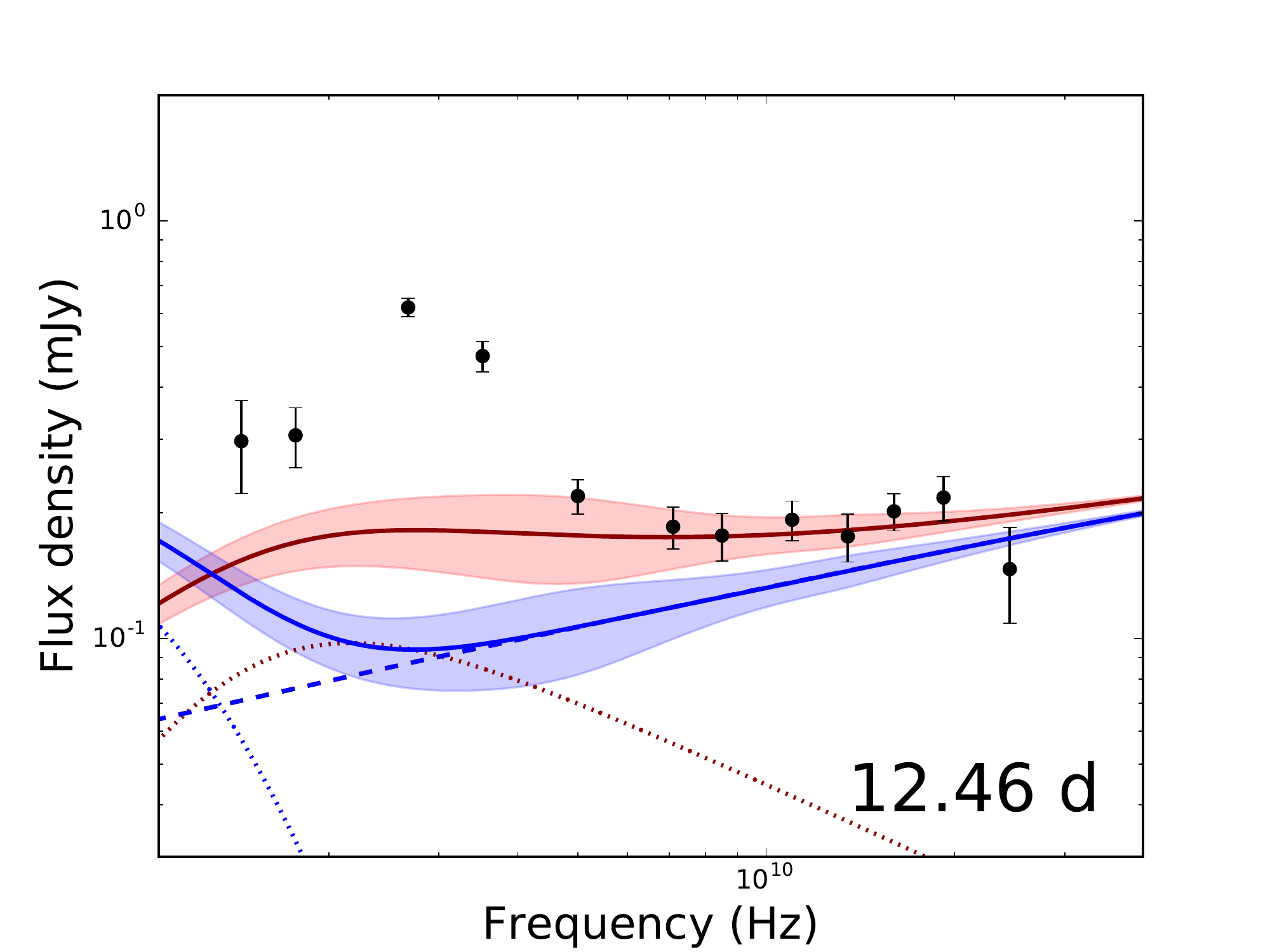}\includegraphics[width=2.5in]{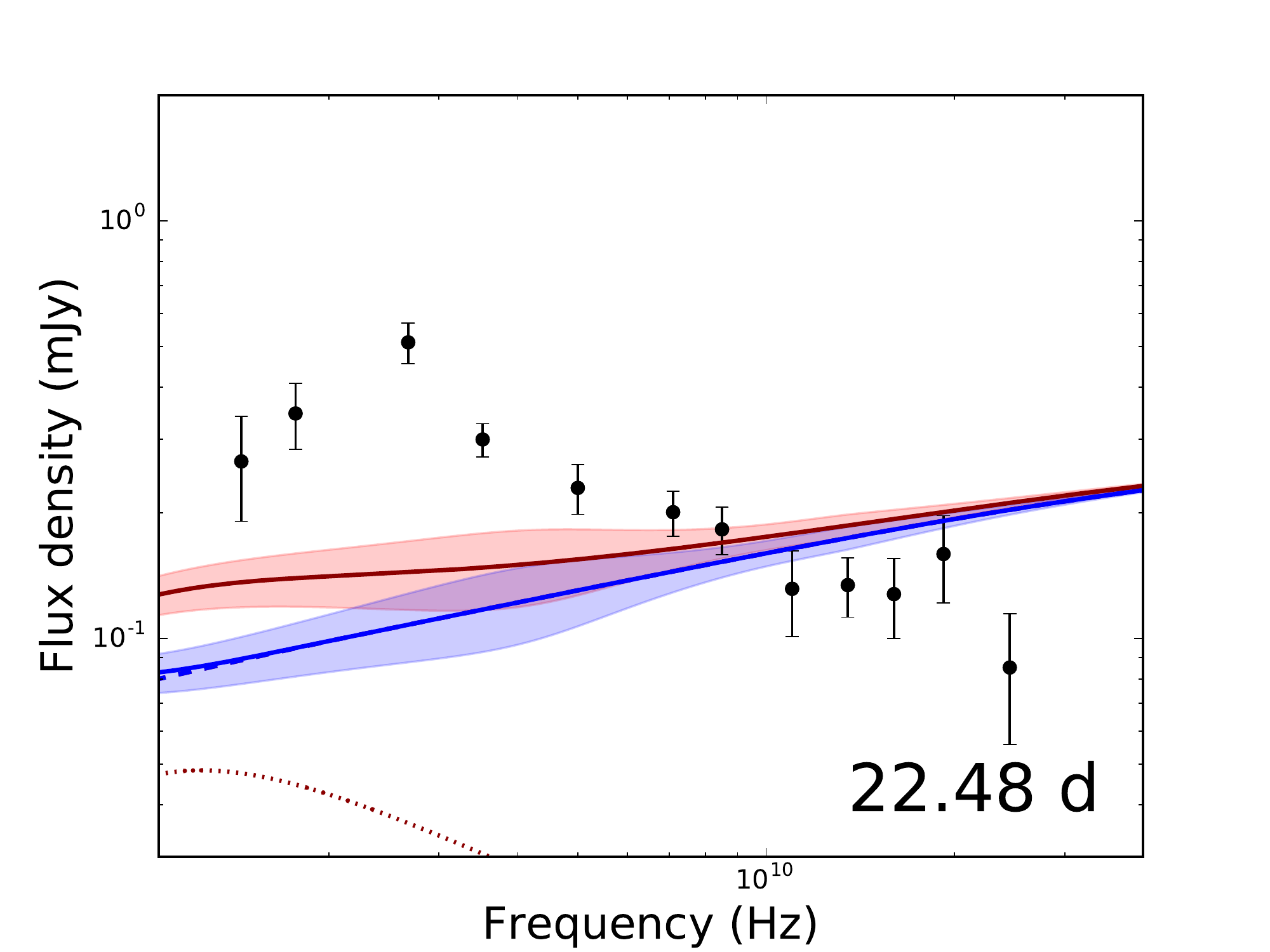}\includegraphics[width=2.5in]{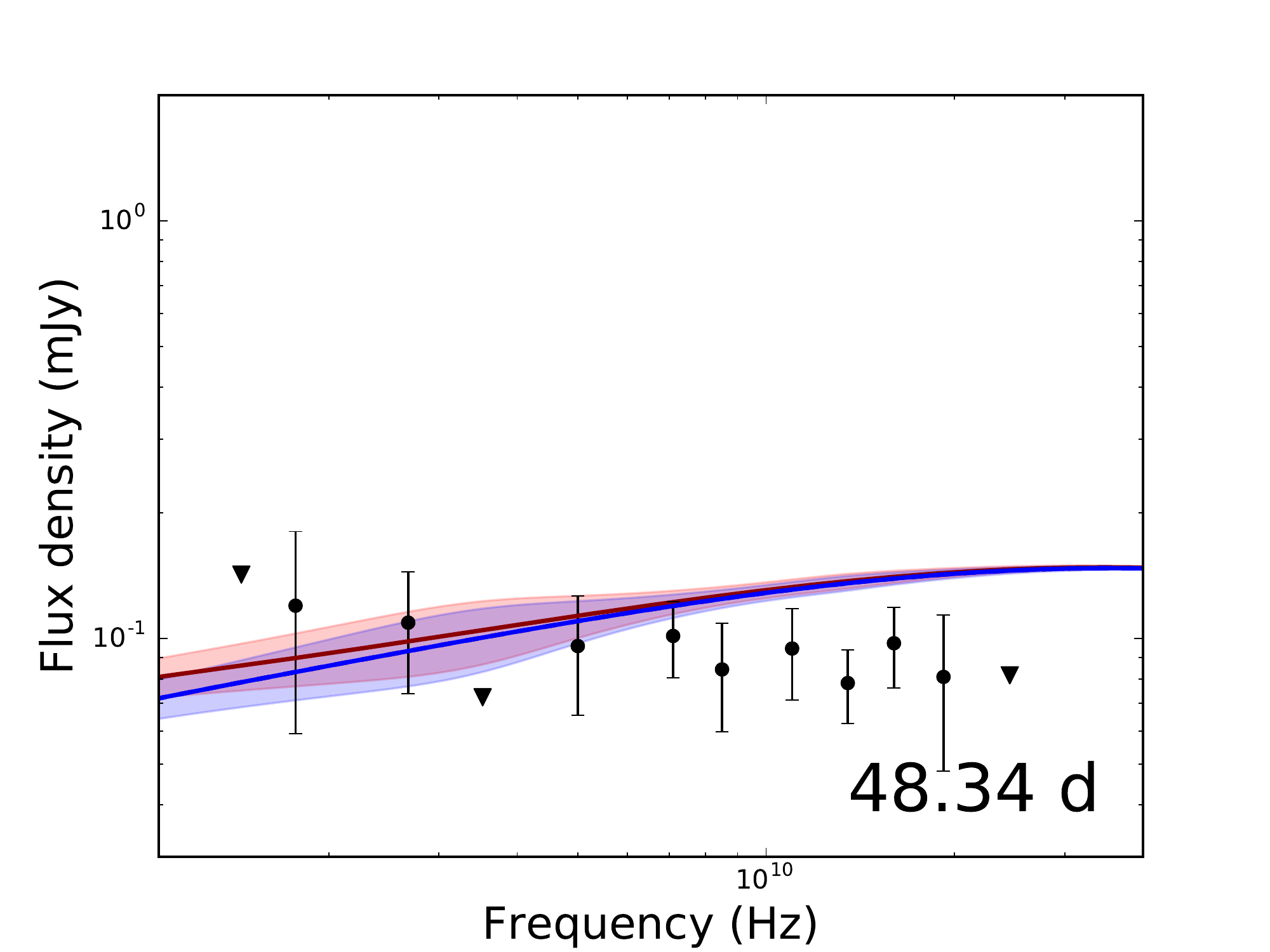}} 
\caption{Observed radio spectral energy distributions of GRB 160625B (black points) with two possible synchrotron models (solid lines) consisting of emission from a forward shock (dashed lines) and a reverse shock (dotted lines). The shaded bands give the expected amplitude of fluctuations caused by interstellar scintillation in the standard thin screen approximation from NE2001 \citep{cl02,gn06}. The FS is the same in both models but we show two different RS models: a Newtonian RS with $\nu_p=\nu_a$, $g=3.5$ and $t_{\rm dec}=400$ s (red; Model 1), and a Newtonian RS with $\nu_p=\nu_c$, $g=1.5$, and $t_{\rm dec}=690$ s (blue; Model 2). The first two epochs are dominated by emission from the RS, while the last epoch is dominated by the FS. The intermediate epochs show the appearance of a third component, whose spectral and temporal evolution cannot be explained in a standard RS + FS model (Section \ref{sec:rad}). The model parameters are given in Table \ref{tab:mod}. Model 2 provides a better fit to epoch 1, but a worse fit to the high frequency data in epochs $2-4$.}
\label{fig:sed2}
\end{figure*}

\begin{figure*} 
\centerline{\includegraphics[width=8in]{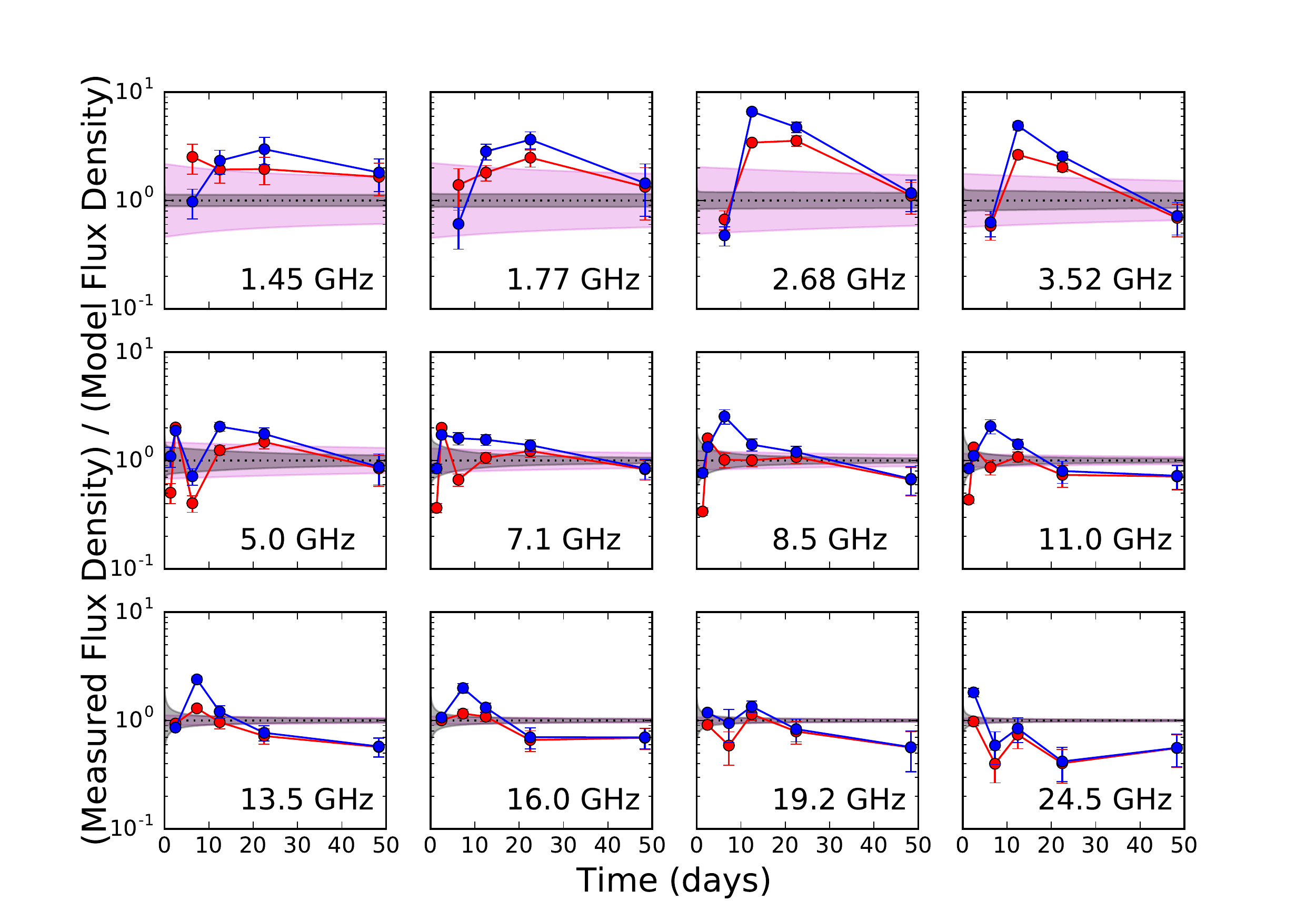}}
\caption{Radio light curves of GRB 160625B constructed by dividing the observed flux density in each band by the FS model given in Table \ref{tab:mod} plus one of two Newtonian RS models. The red points show RS Model 1 and the blue points show RS Model 2 (Table \ref{tab:mod}; Section \ref{sec:rs}). The shaded bands show the \cite{gn06} $1\sigma$ amplitude of ISS fluctuations at each frequency as a function of time using the NE2001 model (gray; $d_{\rm scr}=2.2$ kpc) and a model with $d_{\rm scr}=10$ pc (magenta). The bandwidth of the observations at each frequency is $\sim1$ GHz, except at 1.45 GHz and 1.77 GHz where it is $\sim250$ MHz. The observed variability appears correlated over bandwidths of a few GHz and has an amplitude and duration similar to chromatic ``cusps" previously attributed to plasma lensing of quasars \citep{fied87,fied94,ban16}.}
\label{fig:ESE}
\end{figure*}

We first model the excess radio emission in the early epochs as synchrotron emission from a RS. The RS is launched when the GRB ejecta first begin to interact with the surrounding medium and propagates through the ejecta, probing the properties of the jet itself \citep{sp99,ks00}.  In GRB 160625B, the RS model is constrained by both radio observations and early optical observations. The onset of the optical emission is closely tied to the onset of the main episode of prompt $\gamma$-ray emission: observations by the Mini-MegaTORTORA telescope reveal that the optical flux density increased by a factor of $>90$ in the 30 s prior to the LAT trigger and peaked $\approx12$ s after the LAT trigger time ($\approx3$ s after the $\gamma$-ray peak; \citealt{zzc+16}). This is inconsistent with RS emission models because $T_{90}=35$ s and the RS optical emission is expected to peak at $t_{\rm dec} \geq T_{90}$ \citep{sp99}. We therefore conclude (as do \citealt{llz+17}) that the early optical flash is related to the prompt emission and treat it as an upper bound to the RS emission. 

The RS is most clearly detected in the radio in epochs 1 and 2, so we begin our analysis by fitting this component in these two epochs and then propagate the RS backwards and forwards in time. The radio observations at 1.4 d can be fit with a steeply rising power law with a spectral index $\beta \approx 2$, implying that $\nu_{\rm a,RS} \gtrsim 11$ GHz at this time. Fitting the epoch 2 radio SED with a broken power law, we find that the SED peaks at $\approx 6$ GHz and the spectral index above the peak frequency is $\beta \approx -0.9$. This implies that the peak at 2.5 d is most likely $\nu_a$ (Model 1). In this case, the SED shape also requires $\nu_m \lesssim 6$ GHz and $\nu_c \gtrsim 25$ GHz at 2.5 d. A second possibility is that the peak is $\nu_c$ (Model 2). In this case, $\nu_a \gtrsim 6$ GHz at 2.5 d and $\nu_m$ is unconstrained because the spectrum cuts off above $\nu_c$. This means that various RS models can fit the data equally well, but we show that some models can be ruled out by physical considerations.

\vspace{-0.3in}
\begin{center}
\setlength\LTcapwidth{2in}
\begin{longtable}{lccc}
\label{tab:mod} \\
\caption[]{Model Parameters} \\
\hline
\hline\noalign{\smallskip}
Parameter & Value   \smallskip \\
\hline\noalign{\smallskip} 
\noalign{\center{Forward Shock}\smallskip}
$p$ & $2.31 \pm 0.01$ \\ 
$\epsilon_e$ & $0.23^{+0.07}_{-0.08}$ \\ 
$\log{\epsilon_B}$ & $-1.9^{+1.0}_{-0.9}$ \\ 
$n_0$ & $(5 \pm 3) \times 10^{-5}$ cm$^{-3}$ \\ 
$E_{\rm K, iso}$ & $(1.1^{+1.0}_{-0.5}) \times 10^{54}$ erg \\ 
$t_{\rm jet}$ & $25 \pm 1$ d \\ 
$\theta_{\rm jet}$ & $(3.6 \pm 0.2) ^{\circ}$ \\ 
$E_K\,^a$ & $(2.3^{+1.8}_{-1.2}) \times 10^{51}$ erg \smallskip \\ 
\hline\noalign{\vspace{-0.05in}}
\noalign{\center{Reverse Shock (Model 1)}\smallskip}
$g$ & 3.5 \\
$t_{\rm dec}$ & 400 s \\
$\Gamma_0$ & 290 \\
$R_B$ & 23 \\
$\nu_{a0}$ & $7.88 \times 10^{11}$ Hz    \\      
$\nu_{m0}$ & $6.85 \times 10^{12}$ Hz \\
$\nu_{c0}$ & $2.63 \times 10^{16}$ Hz \\
$f_{\nu_{m0}}$ & 916 mJy \\
\hline\noalign{\vspace{-0.05in}}
\noalign{\center{Reverse Shock (Model 2)}\smallskip}
$g$ & 1.5 \\
$t_{\rm dec}$ & 690 s \\
$\Gamma_0$ & 120 \\
$R_B$ & 630 \\
$\nu_{a0}$ & $8.22 \times 10^{13}$ Hz    \\      
$\nu_{m0}$ & $8.90 \times 10^{13}$ Hz \\
$\nu_{c0}$ & $1.37 \times 10^{14}$ Hz \\
$f_{\nu_{m0}}$ & 2230 mJy \\
\hline\noalign{\vspace{-0.05in}}
\noalign{\center{Reverse Shock (Model 3)}\smallskip}
$g$ & 1.5 \\
$t_{\rm dec}$ & 1300 s \\
$\Gamma_0$ & 370 \\
$R_B$ & 25 \\
$\nu_{a0}$ & $1 \times 10^{12}$ Hz    \\      
$\nu_{m0}$ & $1 \times 10^{12}$ Hz \\
$\nu_{c0}$ & $2 \times 10^{16}$ Hz \\
$f_{\nu_{m0}}$ & 1000 mJy \\
\hline\noalign{\smallskip}
\caption[]{The values given for each RS model are those plotted in Figures \ref{fig:lc}, \ref{fig:sed2}, \ref{fig:ESE}, and \ref{fig:rrs}, but a range of values are possible for each model (Section \ref{sec:rs}). 
\\ $^a$ Corrected for beaming.}
\end{longtable}
\end{center}


\begin{figure*} 
\centerline{\includegraphics[width=4in]{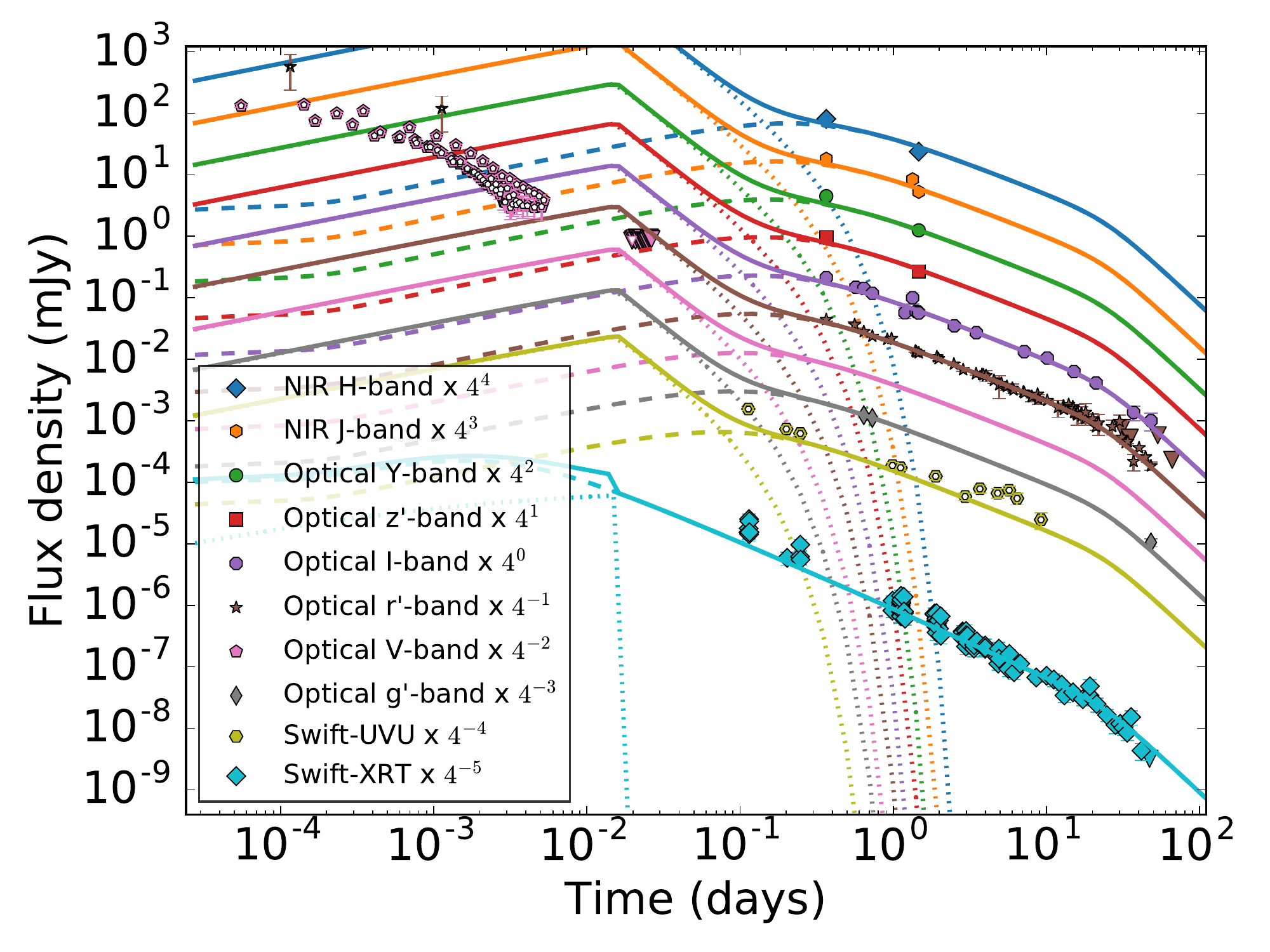}\includegraphics[width=4in]{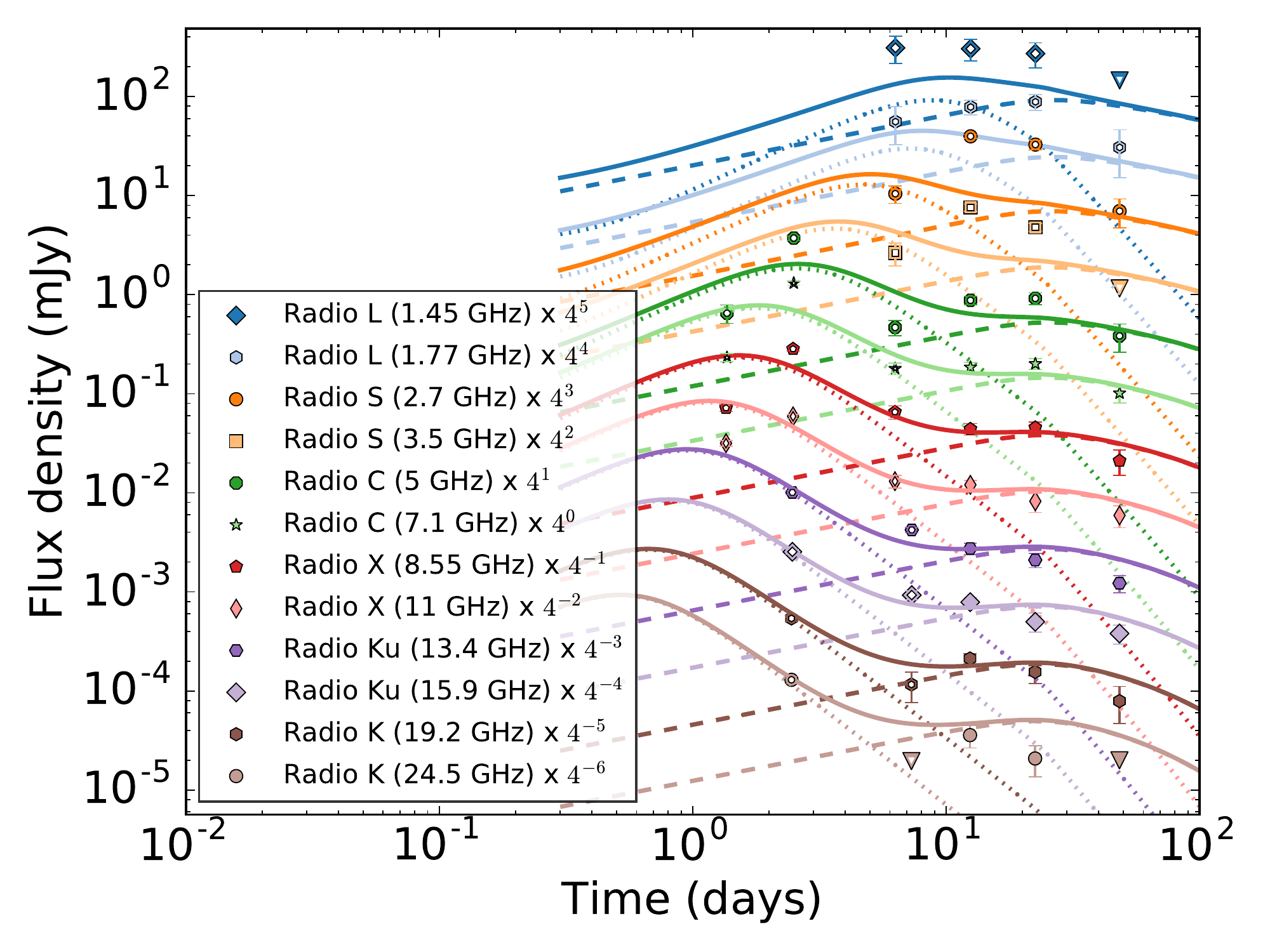}} 
\caption{Same as Figure \ref{fig:lc}, but with a relativistic reverse shock (Model 3; dotted component). The main difference between the Newtonian RS models and the relativistic model shown here is the early optical behavior (left). The fit to the radio data (right) is comparable to Model 1 (Figure \ref{fig:lc}; Section \ref{sec:rs}).}
\label{fig:rrs}
\end{figure*}

In both cases, we run into problems when we attempt to connect the observed SEDs at different epochs. The temporal evolution of the emission depends on whether the RS is relativistic in the frame of the unshocked ejecta. The evolution of the shocked ejecta in a Newtonian RS is characterized by the parameter $g$, which is defined as the rate at which the ejecta Lorentz factor decreases as a function of radius: $\Gamma \propto R^{-g} \propto t^{-g/(1+2g)}$. In the Model 1 case, the best fit to the high-frequency evolution from 2.48 - 12.46 d is obtained for $g\approx3.5$. However, this model does not fit the low-frequency data well for any value of $g$; it overpredicts the emission at 1.36 d and underpredicts the peak at 2.48 d. A perfect fit to the data below 19 GHz at 1.36 d and 2.48 d can be obtained for Model 2 with $g\approx0.2$, but this model would strongly underpredict the emission at all frequencies at 6.8 d and beyond. Furthermore, theoretical constraints limit $g$ to the range $1.5 \leq g \leq 3.5$ for an ISM environment \citep{ks00}; a value of $g<1.5$ would imply that the ejecta has outpaced the FS. The best overall fit for Model 2 is obtained for $g\approx1.5$, which fits the SED at 1.36 d and the low-frequency observations at 6.8 d quite well but underpredicts the high-frequency flux density at 6.8 d. We show the best fits for Model 1 (red) and Model 2 (blue) together with the observed radio SEDs in Figure \ref{fig:sed2}. Neither model reproduces the low-frequency peak in epochs 4 and 5; we return to this point in Section \ref{sec:rad}. The ratio between the observed flux density and the model flux density at each frequency as a function of time is shown in Figure \ref{fig:ESE}. Overall, Model 1 provides a better fit to the data at late times and higher frequencies, where we expect the flux distortions due to propagation effects to be smaller (shaded bands in Figures \ref{fig:sed2} and \ref{fig:ESE}; Section \ref{sec:rad}). 

A similar analysis can be carried out for relativistic RS models. These models are mainly distinguishable from the Newtonian RS models in their predictions for the early optical emission. Relativistic models where the peak frequency is defined by $\nu_c \approx 6$ GHz at 2.5 d are ruled out because they overpredict the observed optical emission $\approx200-300$ s after the burst. Models where $\nu_{\rm m,RS} \lesssim \nu_{\rm a,RS} \approx 6$ GHz predict fluxes much lower than the observed optical fluxes at $t<0.03$ d, again implying that the optical emission originates separately (Model 3; Figure \ref{fig:rrs}). Model 3 and Model 1 produce nearly identical radio SEDs at the times of our observations, so Model 3 is not shown in Figure \ref{fig:ESE}. The exact parameter values chosen for plotting purposes are shown in Table \ref{tab:mod} for each of the three RS models.

Consistency arguments require that the break frequencies of the RS and the FS are related at $t_{\text{dec}}$, the time at which the RS finishes crossing the ejecta. This allows for a measurement of the bulk Lorentz factor ($\Gamma_0$) and the RS magnetization ($R_B\equiv\epsilon_{\text{B,RS}}/\epsilon_{\text{B,FS}}$) at this time \citep{gom08,hk13}. The loose constraints on one or more break frequencies in each model mean that we can only place limits on these quantities, rather than estimate them precisely. In particular, models with shorter $t_{\rm dec}$ values require larger values of $R_B$. For Models 1 and 3, we find $\Gamma_0 \gtrsim 100$ and $1 \lesssim R_B \lesssim 100$, where $\Gamma_0$ is globally minimized for $R_B\approx1$. The relativistic models require slightly longer deceleration times; $t_{\rm dec} \gtrsim120$ s for Model 1, while $t_{\rm dec} \gtrsim480$ s for Model 3. For Model 2, we find $t_{\rm dec} \gtrsim 690$ s and $R_B \gtrsim 630$. Model 2 cannot place any limits on $\Gamma_0$ because $\nu_{\rm m, RS}$ is completely unconstrained in this case.

We can rule out some of these models by requiring $\epsilon_{\rm B, RS} < 1/3$, as we did with $\epsilon_{\rm B, FS}$ in Section \ref{sec:mod}. From the distribution in Figure \ref{fig:hists}, we find that $\epsilon_{\rm B, FS} > 1.56\times10^{-3}$ with 95\% confidence. This requires $R_B < 214$, which is in tension with the lower limit on $R_B$ found for Model 2. For $\epsilon_{\rm B, FS}=0.0136$ (the median of the distribution), we require $R_B < 25$ and the corresponding lower limit on $t_{\rm dec}$ increases, becoming $t_{\rm dec} \gtrsim 400$ s for Model 1 and $t_{\rm dec} \gtrsim 1300$ s for Model 3. The Model 3 limit is problematic because for relativistic RS models we expect $t_{\rm dec} \approx T_{90}$ \citep{kob00}. In GRB 160625B, weak $\gamma$-ray emission was observed until $\sim10$ minutes after the LAT trigger time (Section \ref{sec:he}), but even if we take $T_{90} \approx 600$ s we find that $t_{dec}$ is longer than expected unless $R_B \gtrsim 80$. We therefore conclude that Model 3 is consistent with the data but prefers lower values of $\epsilon_{\rm B, FS}$ than we would predict from the FS modeling alone. If this model is correct, it illustrates how additional information from the RS can break some of the FS parameter degeneracies we found in Section \ref{sec:mod}. A full FS + RS joint MCMC analysis is beyond the scope of this paper and would require better time sampling of the scattering effects discussed in the next section, which currently dominate the RS modeling uncertainties. 

In summary, physical considerations clearly favor Model 1 or 3 over Model 2. Although we cannot distinguish between a relativistic and a Newtonian RS, both models place similar limits on the initial Lorentz factor and the magnetization of the ejecta, $\Gamma_0 \gtrsim 100$ and $1 \lesssim R_B \lesssim 100$. Both models require a deceleration time longer than $T_{90}$ for the main $\gamma$-ray emission episode, slightly disfavoring Model 3 because relativistic RS models predict $t_{\rm dec} \approx T_{90}$. In future events, a joint analysis of well-sampled RS and FS components may enable better constraints on the burst parameters than is possible from observations of either component alone.

\subsection{Late-Time Low-Frequency Rebrightening: An Extreme Scattering Event?}\label{sec:rad}

The late-time radio emission from $12-22$ d is characterized by an abrupt rebrightening centered at 3 GHz that cannot be explained by the fading RS discussed above. Unlike the RS and FS synchrotron emission components, this component is spectrally narrow and only dominates the emission between $1-5$ GHz. Furthermore, the peak flux density $F_{\nu,p}$ and peak frequency $\nu_p$ show unusual time evolution. We parameterize the time evolution of these quantities as $F_{\nu,p} \propto t^a$ and $\nu_p \propto t^b$, but find that the data are inconsistent with single values of $a$ and $b$. Between 12 and 22 days, $F_{\nu,p}\approx0.5$ mJy and $\nu_p\approx3$ GHz remain approximately constant. Before 12 days, the RS dominates the emission so the evolution of these quantities is poorly constrained, but we see that to hide the emission from this component at 7 days either $a$ or $b$ must be nonzero: we require $F_{\nu,p} \lesssim 0.1$ mJy or $\nu_p \gtrsim 25$ GHz, implying $a \gtrsim 3$ or $b \lesssim -4$ from $7-12$ days. The excess vanishes by 48.34 d, implying $F_{\nu,p} \lesssim 0.1$ mJy or $\nu_p \lesssim 1.5$ GHz at this time and requiring $a \lesssim -2$ or $b \lesssim -0.5$ from $22-48$ days. Below, we present several possible explanations for this late-time component, considering both processes intrinsic to the burst and propagation effects that distort the radio spectrum.

\subsubsection{Intrinsic Effects}

We first consider whether an additional synchrotron emission component, such as a second RS, can explain the late-time rebrightening. Like the FS and RS emission discussed above, its SED would consist of smooth power law segments characterized by several break frequencies and an overall normalization. These break frequencies are predicted to evolve in time at constant rates $t^b$, but this is inconsistent with the variable time evolution described above, especially the rapid appearance of this emission component between 7 and 12 d. Furthermore, the narrowness of the emission component leads to spectral indices below and above the peak that are too sharp for standard RS or FS emission (Section \ref{sec:mrc}). 

Some of the problematic time evolution can be avoided if we consider a ``refreshed" RS launched significantly after the prompt emission by the collision of two decelerated shells of ejecta with different initial Lorentz factors \citep{vla11}. The lack of radio emission from this component at $t<12$ days is expected if such a collision happens $\sim10$ days after the GRB, but in such a model we would expect the peak flux and frequency of this component to decrease rapidly at $t>10$ d, inconsistent with the roughly constant flux we observe from $12 - 22$ d. Furthermore, the collision would inject additional energy into the FS, so we would expect to see a late-time plateau or rebrightening at higher frequencies dominated by FS emission. The well-sampled $i^{\prime}$ band, $r^{\prime}$ band, and X-ray light curves show no deviations from smooth power law decline preceding or during the appearance of the late-time radio component (Figure \ref{fig:lc}), so such models are ruled out. We conclude that neither a standard RS nor a ``refreshed" RS can explain this emission.

Variability inconsistent with standard synchrotron afterglow models has been seen in X-ray and optical light curves of long GRBs previously (see \citealt{zha07} for a review). X-ray and optical plateaus, flares, and rebrightenings have been variously attributed to late-time central engine activity, continuous energy injection from ejecta with a range of initial Lorentz factors that collide too gently to produce RS emission, structured jets, variations in microphysical parameters, and deviations of the circumburst density profile from a smooth constant or wind-like profile \citep{pan06,lp07,kong10,uz14,lbm+15,geng16}. Much of this unusual behavior takes place minutes to hours after the burst, rather than tens of days. Furthermore, all of these mechanisms are predicted to produce detectable emission at all frequencies, not just in the radio band, and we see no evidence of a broadband rebrightening in the X-rays or optical on any timescales probed by our observations (Figure \ref{fig:lc}). We conclude that the radio variability we observe in GRB 160625B has a different origin from previously-observed X-ray and optical variability in GRB afterglows.  

To summarize, the late onset, long duration, and highly chromatic nature of the rebrightening are difficult to reconcile with any model in which this component is emission intrinsic to the source. We therefore consider models in which the emitted SED is distorted by propagation effects between the point of emission and the observer.

\subsubsection{Interstellar Scintillation}\label{sec:ISS}

Inhomogeneities in the electron density distribution along the line of sight cause interstellar scintillation (ISS), which distorts radio waves propagating through the Galactic interstellar medium and produces observable flux variations in compact extragalactic radio sources like GRB afterglows and quasars \citep{rick90,good97,walk98,gn06}. ISS is strongly frequency dependent: at high radio frequencies only modest flux variations are expected, while at low frequencies both strong diffractive and refractive effects are important. In the standard picture, all scattering is assumed to occur at a single ``thin screen" located at a distance determined by the NE2001 model for the Galactic electron distribution \citep{cl02}, typically $\sim1$ kpc for high Galactic latitudes. We use this assumption to estimate the transition frequency between strong and weak scattering, $\nu_T\sim 15$ GHz for GRB 160625B. In the strong ISS regime, diffractive scintillation can produce large flux variations on timescales of minutes to hours but is only coherent across a bandwidth $\Delta\nu/\nu=(\nu/\nu_T)^{3.4}$ \citep{good97,walk98}. Since the typical bandwidth of our radio observations is about 1 GHz, we only expect diffractive scintillation to contribute significantly to the observed variability near $\nu_T$. Refractive scintillation is broadband and varies more slowly, on timescales of hours to days. In all regimes, the expected strength of the modulation decreases with time at all frequencies as the size of the emitting region expands, with diffractive ISS quenching before refractive ISS. The source expansion also increases the typical timescale of the variations for both diffractive and refractive ISS.

The shaded bands in Figure \ref{fig:sed2} show the expected strength of ISS in each of our radio epochs based on this simple picture, following \cite{gn06} and including both diffractive and refractive contributions. Clearly, the standard approach cannot explain the full amplitude of the low-frequency peak at 12 d and 22 d, although some of the deviations from the RS models explored in Section \ref{sec:rs} are likely explained by ISS. The large amplitude of this component in the context of ISS suggests diffractive rather than refractive ISS. The spectral width of this feature $\Delta\nu/\nu \sim 1$ and the fact that the variability abruptly cuts off above 3.5 GHz together suggest that $\nu_T\sim 3.5$ GHz (rather than 15 GHz as determined from the NE2001 model). The value of $\nu_T$ is given by $\nu_T \approx 11.6 (d_{\rm scr}/1 \text{ kpc})^{5/17}$ GHz, implying that the scattering screen is located at a distance of $d_{\rm scr}\approx 20$ pc \citep{good97}. The timescale for diffractive ISS at 2.7 GHz is $\approx30$ minutes, much shorter than the $\approx10$ days that the excess endures, but longer than the time on source in each epoch (14 minutes). We see no evidence of variability at 2.7 GHz within a single observation, but the signal-to-noise ratio is low. With only two observations during this time period, it is possible that we caught an upward fluctuation twice.

Since diffractive ISS is only effective for compact sources, we can use the duration of the observed variability to obtain an independent estimate of the size of the emitting region. The maximum angular size for diffractive scintillation at 2.7 GHz is $\theta_s = 94(\nu/10 \text{ GHz})^{6/5}(d_{\rm scr}/{\rm kpc})^{-1} \approx 20$ $\mu$as for a screen distance $d_{\rm scr}=20$ pc \citep{good97}. The strong variability is not present in our final epoch, so we assume that the angular size of the afterglow increased past $\theta_s$ sometime between 22 d and 48 d. Our FS model predicts that the angular size of the afterglow is 40 $\mu$as at 22 d and 60 $\mu$as at 48 d, which is consistent with this limit to within a factor of two. Exactly matching the FS prediction would require a slightly closer screen at $\approx7-10$ pc, which is also roughly consistent with the low-frequency observations. In Figure \ref{fig:ESE}, we show the predicted $1\sigma$ variations due to ISS for $d_{\rm scr}=10$ pc (magenta shaded region) and the standard NE2001 prediction $d_{\rm scr}=2.2$ kpc (gray shaded region). The $d_{\rm scr}=10$ pc model does a better job of explaining the variability at frequencies below 5 GHz, but underpredicts the observed variations at 7-11 GHz in epoch 1. Both models fail to reproduce the late-time flux deficit at high frequencies noted in Section \ref{sec:tjet}, although many of these points have large error bars due to the faintness of the fading afterglow.

GRB 160625B is not the first source in which non-standard ISS models have been invoked to explain extreme variability. An even closer scattering screen ($d_{\rm scr}=1-2$ pc) was previously inferred for the quasar J1819+3845, which showed extreme variability that stopped abruptly after 7.5 years and did not return in a further 6 years of monitoring \citep{dm15}. The limited duration of the J1819+3845 variability suggests that the scattering screen was compact or patchy, which may also be the case for the nearby structure responsible for the strong flux modulations we see in GRB 160625B. We note that the extreme amplitude, bandwidth, and duration of this component are also qualitatively similar to extreme scattering events (ESEs) observed in quasars \citep{fied87,fied94,ban16}. While ISS has been observed in other GRB afterglows (e.g. \citealt{wkf98,bkp+03,chan08}), this would make GRB 160625B the first GRB to exhibit an ESE. The proposed cause of ESEs is lensing by dense $\sim$ AU-scale plasma structures in the Milky Way that transit the line of sight. Such structures are not dissimilar to the $\sim100$ AU-scale object proposed as the cause of the extreme variability in J1819+3845 \citep{dm15}. As with the \cite{fied94} ESEs, the variability in GRB 160625B is uncorrelated across bandwidths larger than a few GHz (Figure \ref{fig:ESE}). In other literature ESEs, a rapid flux enhancement is followed by an extended period ($\sim$ months) in which the flux is suppressed and then by a second enhancement, producing chromatic symmetric U-shaped features. The amplitude ($\approx3$ times the predicted model flux) and duration ($\approx10$ days) of the 2.7 GHz feature are comparable to the flux enhancements seen during these bracketing cusps. A search for long-lasting flux suppression before or after the observed enhancement is complicated by uncertainties in the afterglow modeling, limited wavelength coverage before 6 d, increased flux uncertainties at later times due to the fading of the afterglow, and the more sparse time sampling after 12 d. We note that the rapid flux variations at $1-6$ d at $5-9$ GHz are somewhat reminiscent of the sharp features observed at 8.5 GHz in an ESE towards the quasar 0954+658 during the 2.7 GHz event minimum \citep{fied87}, which would mean that the observed flux increase in GRB 160625B corresponds to the end of the proposed ESE. 

We conclude that the excess low-frequency emission observed in GRB 160625B from $12-22$ d is broadly consistent with previously observed variability in compact extragalactic sources attributed to diffractive ISS or other extreme scattering effects. The observations suggest that much of the scattering occurs at a distance of $\approx 10-20$ pc, much closer than is typically assumed. A combination of scattering from this nearby screen and the more distant ``standard" screen could explain the additional variability observed at $1.4-6.8$ d. Future GRB observations with broad frequency coverage and denser time sampling will better constrain the timescales of such variability and allow us to disentangle ISS from variations intrinsic to the source.

\section{Conclusions}

We presented detailed observations of the long-duration GRB 160625B spanning radio to X-ray wavelengths and found that the data are mostly well-fit by the standard forward shock model for GRB afterglows. We use a MCMC analysis to constrain the afterglow properties and find that GRB 160625B is a highly energetic event that exploded in an ISM-like low-density medium. Our early multi-frequency radio observations show a clear excess compared to the standard predictions for synchrotron emission from a forward shock. We interpret this excess as a reverse shock, making GRB 160625B only the third GRB for which an in-depth study of RS emission at multiple epochs has been possible. All three events occurred in low density environments, suggesting that such conditions are particularly favorable for the production of strong, long-lasting RS emission. Our ability to constrain the jet properties is restricted by the limited wavelength coverage of our first epoch and by the additional uncertainty introduced by interstellar scintillation, which causes large random flux perturbations at low frequencies in our first five radio epochs. We place a lower limit on the initial bulk Lorentz factor of the ejecta of $\Gamma_0 \gtrsim 100$ that is robust to other uncertainties in the RS modeling, confirming the highly-relativistic nature of the outflow. The magnetization of the RS is $R_B \approx 1-100$. 

One key finding from this analysis is that propagation effects cannot be ignored when attempting detailed physical characterization of GRB radio afterglows, especially at early times when RS emission is most relevant. The radio afterglow of GRB 160625B shows unusual variability on a range of timescales, most notably a low-frequency rebrightening centered at 3 GHz at $12-22$ days. This late-time excess cannot be easily explained with processes intrinsic to the source. Instead, it is more naturally explained in the context of propagation effects in the Galactic ISM, and is roughly consistent with strong diffractive scintillation by a thin screen with an effective distance of $\approx10-20$ pc. The extreme variability at 2.7 GHz is qualitatively similar to plasma lensing by compact structures in the Milky Way. A more detailed analysis of this intriguing similarity is not possible for GRB 160625B because our observing strategy, while a significant improvement on previous efforts, is optimized to probe RS emission at early times rather than more rapid ISS-induced variability that may endure for several weeks. Disentangling propagation and intrinsic effects will require denser time and frequency coverage of GRB radio afterglows than has been attempted to date, but will enable new probes of both GRB physics and the nature of turbulent structures in the ISM. We will further explore the impact of propagation effects on GRB afterglows in future work.

\begin{acknowledgements} We thank R.~Barniol Duran, M.~Johnson, R.~Narayan, R.~Sari, D.~Warren, B.-B.~Zhang, and the attendees of the Eighth Huntsville Gamma-Ray Burst Symposium for useful conversations. We also thank the anonymous referee for helpful comments that have improved this manuscript. K.D.A.~and E.B.~acknowledge support from NSF grant AST-1411763 and NASA  ADA   grant NNX15AE50G. T.L. is a Jansky Fellow of the National Radio Astronomy Observatory (NRAO). W.F. is supported by NASA through Einstein Postdoctoral Fellowship grant number PF4-150121. VLA observations were taken as part of our VLA Large Program 15A-235 (PI: E. Berger). The VLA is operated by the NRAO, a facility of the National Science Foundation operated under cooperative agreement by Associated Universities, Inc. This work made use of data supplied by the UK {\it Swift} Science Data Centre at the University of Leicester. \end{acknowledgements}

\software{CASA \citep{casa}, pwkit \citep{pwkit}} 

\bibliographystyle{aasjournal}
\bibliography{160625B_v5}

\end{document}